\documentclass[sigconf]{acmart}

\copyrightyear{2026}
\acmYear{2026}
\setcopyright{cc}
\setcctype{by}
\acmConference[KDD 2026] {Proceedings of the 32nd ACM SIGKDD Conference on Knowledge Discovery and Data Mining V.2}{August 9--13, 2026}{Jeju Island, Republic of Korea.}
\acmBooktitle{Proceedings of the 32nd ACM SIGKDD Conference on Knowledge Discovery and Data Mining V.2 (KDD 2026), August 9--13, 2026, Jeju Island, Republic of Korea}
\acmISBN{979-8-4007-2259-2/2026/08}
\acmDOI{10.1145/3770855.3818827}

\AtBeginDocument{%
  }

\usepackage{graphicx}
\usepackage{xcolor}         
\usepackage{amsmath}
\usepackage{multirow}    
\usepackage{pifont}      
\usepackage{tablefootnote} 
\usepackage{enumitem}
\usepackage{subfigure}
\usepackage{wrapfig}  

\usepackage{colortbl}  
\definecolor{gray94}{gray}{.94}
\definecolor{gray90}{gray}{.90}
\newcommand{\grow}[1]{\rowcolor{gray94}{#1}}
\begin{document}

\title{SurfDesign: Effective Protein Design on Molecular Surfaces}
\author{Fang Wu}
\orcid{0000-0001-7240-3915}
\affiliation{\institution{Stanford University}\city{Stanford}\country{USA}}
\email{fangwu97@stanford.edu}

\author{Shuting Jin}
\authornote{Corresponding author.}
\orcid{0000-0002-8113-9367}
\affiliation{\institution{Wuhan University of Science and Technology}\city{Wuhan}\country{China}}
\email{shutingjin@wust.edu.cn}

\author{Xiangru Tang}
\orcid{0009-0006-2700-4513}
\affiliation{\institution{Yale University}\city{New Haven}\country{USA}}
\email{xiangru.tang@yale.edu}

\author{Mark Gerstein}
\orcid{0000-0002-9746-3719}
\affiliation{\institution{School of Medicine, Yale University}\city{New Haven}\country{USA}}
\email{pi@gersteinlab.org}

\author{Xiangxiang Zeng}
\orcid{0000-0003-1081-7658}
\affiliation{\institution{Hunan University}\city{Changsha}\country{China}}
\affiliation{%
  \institution{Yuelushan Laboratory}
  \city{Changsha}
  \country{China}
}
\email{xzeng@hnu.edu.cn}

\author{Jure Leskovec}
\orcid{0000-0002-5411-923X}
\affiliation{%
  \institution{Kumo.AI}
  \city{Mountain View}
  \country{USA}
}
\affiliation{\institution{Stanford University}\city{Stanford}\country{USA}}
\email{jure@cs.stanford.edu}

\author{Yejin Choi}
\orcid{0000-0003-3032-5378}
\affiliation{\institution{Stanford University}\city{Stanford}\country{USA}}
\email{yejinc@cs.stanford.edu}

\author{Jinbo Xu}
\orcid{0000-0001-7111-4839}
\affiliation{\institution{Toyota Technological Institute at Chicago}\city{Illinois}\country{USA}}
\email{jinboxu@gmail.com}


\renewcommand{\shortauthors}{Fang et al.}

\begin{abstract}
    Protein function is largely determined by molecular surface geometry and physicochemical complementarity, yet most protein design methods condition only on backbone structure. We introduce SurfDesign, a surface-conditioned protein design framework that models molecular surfaces as continuous geometric manifolds and integrates them with pretrained protein language models. SurfDesign employs surface-based equivariant message passing to capture surface normals, curvature, and directional geometry, together with a parameter-efficient fine-tuning strategy. Focusing on functional protein design, we show that SurfDesign consistently outperforms prior surface-conditioned and backbone-only methods on de novo binder and enzyme design benchmarks. We also report strong performance on inverse-folding benchmarks as a diagnostic of structural compatibility. Our results highlight manifold-aware surface representations as a principled foundation for functional protein and enzyme design. Code is available at~\url{https://github.com/smiles724/SurfDesign}
\end{abstract}

 \ccsdesc[500]{Applied computing
  -> Life and medical sciences
       -> Computational biology}

\keywords{Protein Design, Surface Modeling}



\maketitle

\section{Introduction}

Proteins are fundamental molecular machines that drive nearly all biological processes, including catalysis, molecular recognition, signaling, and regulation. Recent advances in deep learning (DL) have substantially accelerated progress in protein design, shifting the field from physics-based optimization toward data-driven generative modeling~\citep{ingraham2019generative,jing2020learning,wang2024diffusion,wu2024hierarchical,wu2024semi,wu2026semi,wudiffantiseq,wu2022pre}. A dominant paradigm in this space is structure-based inverse folding~\citep{defresne2021protein}, where a target backbone is specified, and a compatible amino-acid sequence is generated. A large body of work has demonstrated impressive improvements under this formulation, especially with graph neural networks (GNNs) and pretrained protein language models (PLMs).
\begin{figure}[t]
\begin{center}
\includegraphics[width=\columnwidth]{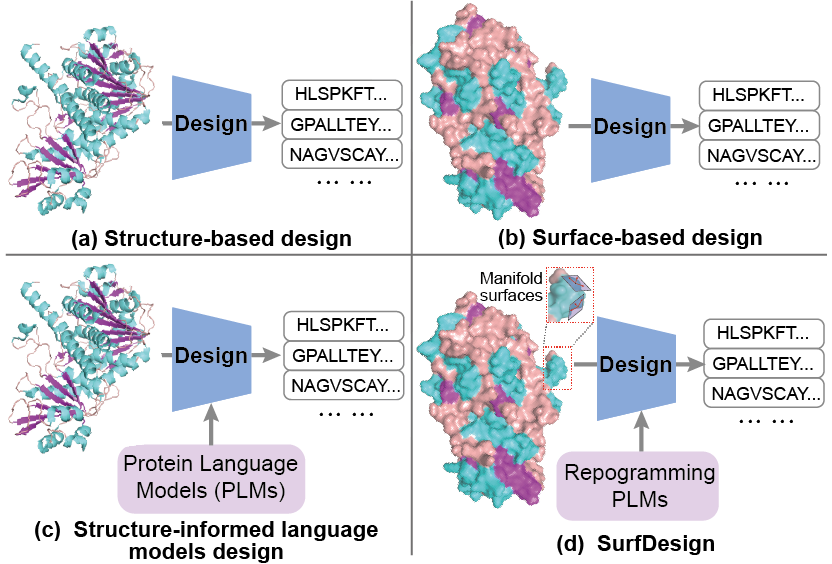}
  \end{center}
  \caption{Protein design setups, conditioned on backbone structures or molecular surfaces. }
  \vspace{-2em}
  \label{fig:protein_design}
\end{figure}

However, the ultimate objective of protein design extends beyond folding correctness~\citep{song2024surfpro,tang2025bc}. Many practical design tasks, such as enzyme engineering, receptor-ligand binding, and de novo interaction design, are governed not solely by backbone geometry, but by localized surface shape and physicochemical complementarity~\citep{wu2024surface,wu2025surface,li2025joint}. Proteins with nearly identical folds can exhibit drastically different functions if their surface charge distributions, curvature, or hydrophobic patterns differ. Consequently, backbone-only approaches may fail to adequately constrain functional interfaces, even when global folding is correct~\citep{li2025surffold}.

Molecular surfaces provide a natural and functionally grounded representation. Defined by smooth atomic boundaries, protein surfaces directly mediate physical interactions with substrates, ligands, and binding partners. Surface geometry encodes shape complementarity, while surface-level physicochemical patterns determine specificity and affinity. Prior work~\citep{song2024surfpro,li2025joint} in protein surface modeling has demonstrated the effectiveness of surface representations for interaction prediction and binder design, highlighting their importance in functional protein discovery. Despite this, surface-conditioned generation remains underexplored, and existing methods often rely on discretized point clouds or meshes that inadequately capture intrinsic continuity and geometry.

Two key challenges exist in surface-conditioned design. First, molecular surfaces are continuous and smooth manifolds~\citep{lee2023pre,sun2024dsr}, whereas existing methods~\citep{song2024surfpro,zhang2023equipocket} treat them as unordered collections of points or fixed meshes, ignoring local tangent structure, curvature, and directional consistency. This mismatch restricts the expressiveness of learned representations and hampers generalization to fine-grained functional regions such as binding pockets and catalytic sites. Second, crystal structures are limited in number relative to available sequences, making it difficult to train data-hungry models purely from surface-sequence pairs. Surface information alone may be insufficient in buried regions, where evolutionary and sequential priors play complementary roles.

\begin{figure*}[t]
    \centering
    \includegraphics[width=\linewidth]{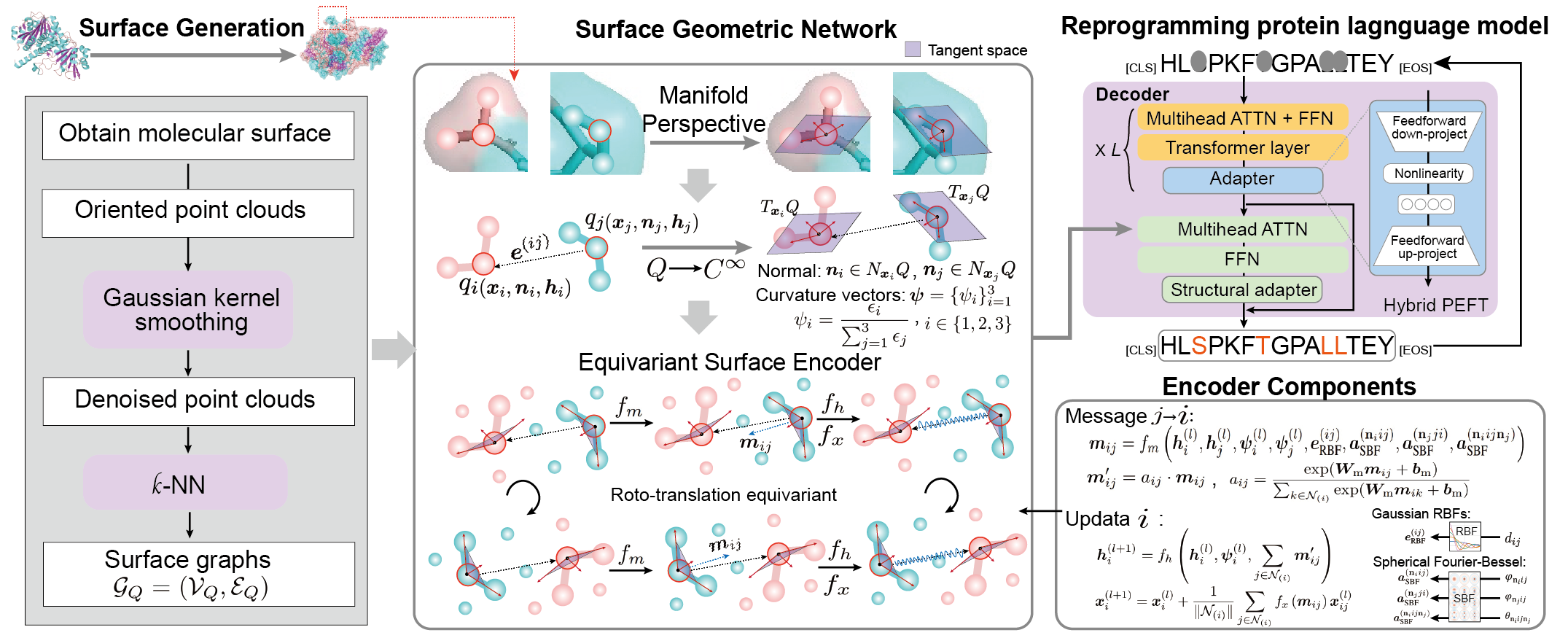}
    \caption{Illustration of SurfDesign. Smooth-surface graphs are obtained using PyMOL or MSMS and subsequently denoised. Then, an equivariant surface encoder is appended to extract manifold representations. These features are further incorporated into the structural adapter of protein language models to recover masked amino acids.  }
    \vspace{-1em}
    \label{fig:model}
\end{figure*}

To address them, we propose SurfDesign that explicitly models molecular surfaces as geometric manifolds and integrates them with PLMs~\citep{qiu2024instructplm,wang2024diffusion} (see Fig.~\ref{fig:model}). We introduce a surface-conditioned equivariant message passing (SEMP) encoder that leverages surface normals, curvature, and directional relationships to capture local manifold structure while preserving roto-translation equivariance. To mitigate data scarcity and enhance sequence modeling, we further incorporate a hybrid parameter-efficient fine-tuning (PEFT) strategy that injects surface-derived structural information into PLMs without full retraining.
Importantly, SurfDesign is designed primarily to generate functional proteins. In functional design tasks such as binding and enzyme recognition, there is no unique ground-truth sequence associated with a given structure or surface. Instead, multiple diverse sequences may satisfy the same functional constraints. Accordingly, we treat inverse folding benchmarks not as supervised label-recovery tasks, but as diagnostic evaluations that assess whether generated sequences are structurally compatible with the specified geometry. This distinction is crucial: \textbf{conditioning on molecular surfaces introduces richer physical constraints, not privileged access to native sequence identities.}

We evaluate SurfDesign on inverse folding benchmarks~\citep{orengo1997cath} and achieve state-of-the-art amino-acid recovery (AAR) and perplexity, validating its ability to generate structurally consistent sequences. More importantly, we show that SurfDesign excels in functional protein design, achieving superior performance in de novo binder generation across six benchmark targets and in enzyme design for multiple enzyme-substrate systems. These findings indicate that expressive surface modeling enables protein design with greater functional fidelity than backbone-only approaches. Overall, this work advances protein design by elevating molecular surfaces from auxiliary features to first-class conditioning signals, demonstrating that explicitly modeling molecular surfaces as geometric manifolds is a powerful and complementary approach for moving beyond backbone-centric design.

\section{Method}
\subsection{Preliminary and Background}
\paragraph{Problem Statement.} Neural structure-conditioned protein design aims to find the amino acid sequence $\mathcal{S}= \{s_i\in \text{Cat}(20): 1 \leq i\leq n \}$ folding into a desired structure  $\mathcal{X} = \{\boldsymbol{x}_i\in \mathbb{R}^{4\times 3}: 1 \leq i\leq n \}$, where $s_i$ is one of the 20 residue types and $\mathcal{X}$ denotes the spatial coordinates for 4 backbone atoms (\emph{i.e.}, $C_\alpha$, $C$, $N$ and $O$). It can be formulated as an end-to-end graph-to-sequence learning problem with a parameterized encoder-decoder neural network $\mathcal{F}_{\boldsymbol{\vartheta}}$: $\mathcal{X} \rightarrow \mathcal{S}$. Surface-conditioned design is analogous but yields functional proteins that fold into the expected surface $\mathcal{Q}$ with associated biochemical properties~\citep{song2024surfpro}. Our objective, therefore, transfers to learn a function $\mathcal{F}_{\boldsymbol{\vartheta}}(\cdot): \mathcal{Q} \rightarrow \mathcal{S}.$ Given sufficient surface-sequence paired data, the learning purpose is to maximize the conditional log-likelihood $p(\mathcal{S} | \mathcal{Q}; \boldsymbol{\vartheta})$. This enables the design of sequences that either maximize likelihood or are generated via sampling algorithms to ensure diversity and novelty. Remarkably, homologous proteins consistently share similar surfaces~\citep{pearson2005limits}, so the surface-conditioned design is underdetermined. 

Unlike supervised classification, protein design does not assume a unique ground-truth sequence  $\mathcal{S}$ for a given structure~\citep{gao2022pifold}. A single backbone or surface geometry is compatible with many valid sequences, and the experimentally observed sequence is neither unique nor necessarily optimal. Accordingly, sequence recovery is used only as a diagnostic proxy for structural compatibility rather than a supervised learning target. Our objective is therefore to learn physically and functionally consistent sequence distributions conditioned on geometric constraints, not to recover native labels. 

\paragraph{Surface Generation}
Surface geometry is crucial for interaction analysis. We employ PyMol~\citep{delano2002pymol} to obtain the raw molecular surface, where a probe of a certain radius ($\sim 1$ \r{A}) is moved along the protein to calculate the Solvent Accessibility Surface (SAS) and Solvent Excluded Surface (SES). We treat the resulting surface vertices as oriented surface points, defined by an oriented point cloud $Q = \{q_i: 1\leq i\leq m\}$ and $m>>n$. Each surface point $q_i$ has a triplet of attributes $(\boldsymbol{x}_i, \boldsymbol{n}_i, \boldsymbol{h}_i)$, where $\boldsymbol{x}_i\in\mathbb{R}^3$ and $\boldsymbol{n}_i\in\mathbb{R}^3$ are the 3D coordinates and unit normal vector, and $\boldsymbol{h}_i\in\mathbb{R}^{\phi_h}$ indicates the physicochemical properties of $q_i$ such as hydrophobicity, hbond, and charge. Then the surface graph is built via $k$-NN, resulting in $\mathcal{G}_Q = (\mathcal{V}_Q, \mathcal{E}_Q)$. We also investigate MSMS~\citep{robinson2014msm} and BioPython~\citep{cock2009biopython} for surface generation and identify negligible differences in processing speeds across several toolkits. As raw point clouds generally carry noise that may limit the expressivity of molecular surfaces~\citep{alexa2001point}, we apply the Gaussian kernel smoothing~\citep{song2024surfpro} to raw cloud data:
\begin{equation}
    \boldsymbol{x}_i \leftarrow \sum_{j\in\mathcal{N}_{(i)}}\frac{\mathcal{K}(\boldsymbol{x}_i, \boldsymbol{x}_j) \cdot \boldsymbol{x}_j}{\sum_{t\in\mathcal{N}_{(i)}} \mathcal{K}(\boldsymbol{x}_i, \boldsymbol{x}_t)}, \quad \mathcal{K}(\boldsymbol{x}, \boldsymbol{y}) =\exp^{-\frac{\|\boldsymbol{x} - \boldsymbol{y}\|^2}{\eta}}, 
\end{equation}
where $\mathcal{N}_{(i)}$ denotes the neighborhood of $\boldsymbol{x}_i$ and $\mathcal{K}(., .)$ is the Gaussian kernel with $\eta$ indicating distance scale in the point space. Here, $\eta$ is set as $\max_{j\in \mathcal{N}_{(i)},i\in[m]}(\left\| \boldsymbol{x}_i - \boldsymbol{x}_j\|^2\right)$. 
For clarity, $\boldsymbol{h}_i$ are computed from local atomic geometry and element types, not from residue identity labels. We do not use: (i) residue identities, (ii) multiple sequence alignments (MSA) or evolutionary profiles, or (iii) functional annotations or active-site labels, during the surface generation pipeline. While we compute surfaces from full-atom structures, these structures may be experimental, predicted, or generated  (e.g., RFdiffusion + packing). The atom types specify geometric constraints but do not reveal the target sequence, as multiple valid sequences can fold into similar atomic arrangements.

\subsection{Surface Geometric Network}
\paragraph{A Manifold Perspective for Molecular Surfaces.} Theoretically, molecular surfaces are continuous manifolds with infinite resolution~\citep{lee2023pre}, which cannot be fully expressed by existing mesh-~\citep{gainza2020deciphering} or point-based~\citep{sverrisson2021fast,zhang2023equipocket,song2024surfpro} mechanisms. 
The key distinguishing property of manifold surfaces relative to conventional point clouds or meshes is that every point on the manifold is locally Euclidean. Mathematically, for $\forall q_i\in Q$, there exists a neighborhood $U_{q_i}$ and a homeomorphism $f_{\text{homo}}(\cdot)$ such that $f_{\text{homo}}: U_{q_i}\rightarrow V\subseteq \mathbb{R}^3$, where $V$ is an open ball in $\mathbb{R}^3$. 
In order to describe the local geometry of a manifold point $q_i\in Q$, we need to know at least (1) the linear approximation of the manifold in its vicinity, which corresponds to \emph{the tangent space}, and (2) how fast the surface bends or deviates from being a plane near this point, which can be measured by \emph{curvature}.  

Towards this goal, we assume that the surface $Q$ is a $C^\infty$ differentiable manifold and $T_{\boldsymbol{x}_i}Q$ denotes the tangent space of any point $\boldsymbol{x}_i\in Q$. Then we can acquire the unit normal vector $\boldsymbol{n}_i\in N_{\boldsymbol{x}_i}Q$ perpendicular to $T_{\boldsymbol{x}_i}Q$. If $Q$ is implicitly described by a signed distance function (SDF) satisfying $f_{\text{SDF}}(\cdot)=0$, then the normal at point $\boldsymbol{x}_i$ is equivalent to the gradient, \emph{i.e.}, $\boldsymbol{n}_i=\nabla f_{\text{SDF}}(\boldsymbol{x}_i)$. Here, we draw the normal vector set $\{\boldsymbol{n}\}_{i=1}^m$ immediately from the software (\emph{i.e.}, PyMol) and integrate this orientation knowledge into the geometric encoder to linearly approximate the manifold and achieve manifold-awareness. Prior studies~\citep{zhang2023equipocket,song2024surfpro} have seldom considered this specialty of molecular surfaces and merely handle naive clouds. One exception, dMaSIF~\citep{strokach2020fast}, notices this manifold uniqueness and computes the quasi-geodesic distance as $d_{ij} = \| \boldsymbol{x}_{ij}\|^2 \cdot \left(2- \boldsymbol{n}_i^\top \cdot \boldsymbol{n}_j\right)$ to naively resemble the geodesic coordinates in the tangent space $T_{\boldsymbol{x}_i}Q$. However, its construction of tangent vectors destroys the equivariance. 

Additionally, there are varying ways to define curvatures of 3D Riemannian manifolds intrinsically without reference to a larger space~\citep{kobayashi1996foundations}, such as normal curvature $k_n$, geodesic curvature $k_g$, and geodesic torsion $\tau_r$. Those all relate the direction of curvatures to the unit normal vector $\boldsymbol{n}_i$. Given a non-singular curve $\gamma(q_i)\in Q$ parametrized by arc length, we can compute $\boldsymbol{T}_i = \gamma'(q_i) = \frac{d\gamma}{dq}$ and $\boldsymbol{t}_i=\boldsymbol{n}_i \times \boldsymbol{T}_i$ to form the Darboux frame. The triple $(\boldsymbol{T}_i, \boldsymbol{t}_i, \boldsymbol{n}_i)$ defines a positively oriented orthonormal basis attached to each point of the curve $\gamma(q_i)$. Then the above quantities are related by $\begin{pmatrix} \boldsymbol{T}' \\ \boldsymbol{t}' \\ \boldsymbol{u}' \end{pmatrix} = \begin{pmatrix} 0 & k_g & k_n\\-k_g & 0 & \tau_r \\-k_n & -\tau_r & 0 \end{pmatrix}\begin{pmatrix} \boldsymbol{T} \\ \boldsymbol{t} \\ \boldsymbol{u} \end{pmatrix}$. Inspired by progress in geometry processing~\citep{tian2023geomae,wu2024surface,zhang2008curvature}, we estimate these quantities in a closed form from local points $\mathcal{N}_{(i)}$. Specifically, we first compute a covariance matrix for $q_i$ and its neighborhood $\mathcal{N}_{(i)}$:
\begin{equation}
\label{equ:covariance}
    \boldsymbol{\Sigma} = \frac{1}{\|\mathcal{N}_{(i)}\|}\sum_{\mathbf{x}_{j}\in \mathcal{N}_{(i)}}\mathbf{x}_{j} \mathbf{x}_{j}^\top - \bar{\mathbf{x}}\bar{\mathbf{x}}^\top,  \quad \boldsymbol{\Sigma}\in \mathbb{R}^{3\times 3}.
\end{equation}
where $\bar{\mathbf{x}}$ is the centroid of this point cluster. After the eigen-decomposition of $\boldsymbol{\Sigma}$ (\emph{e.g.}, singular value decomposition or eigenvalue decomposition), eigenvalues can be obtained as $\epsilon_1, \epsilon_2$, and $ \epsilon_3$ ($\epsilon_1\geq\epsilon_2 \geq\epsilon_3$). Those pseudo curvatures vectors $\boldsymbol{\psi} = \{\psi_i\}_{i=1}^3$ can be therefore computed as:
\begin{equation}
   \psi_i = \frac{\epsilon_i}{\sum_{j=1}^3 \epsilon_j},\quad i \in \{1, 2, 3\}.
\end{equation}
We employ $\boldsymbol{\psi}$ as a rotation-invariant local shape descriptor that approximates curvature-related information, rather than exact differential curvatures. It can be proved that this curvature feature $\boldsymbol{\psi}$ is roto-translation invariant (see App.~\ref{app:curvature}). 

\begin{wrapfigure}{r}{0.4\columnwidth}
\vspace{-1.5em}
  \begin{center}
  \includegraphics[width=0.35\columnwidth]{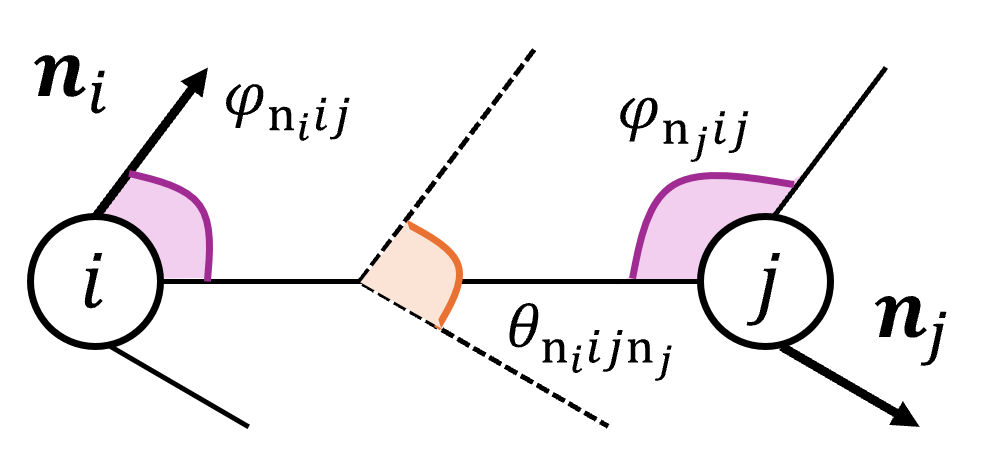}
  \end{center}
  \caption{Angles hidden in the oriented surface point cloud, containing two intersection angles $\varphi_{\mathbf{n}_i ij}=\angle \boldsymbol{n}_i \boldsymbol{x}_{ij}$ and $\varphi_{\mathbf{n}_j ji}=\angle \boldsymbol{n}_{j} \boldsymbol{x}_{ji}$ as well as a dihedral angle $\theta_{\mathbf{n}_i ij \mathbf{n}_j}$.}
  \label{fig:angle}
  \vspace{-1em}
\end{wrapfigure}

\paragraph{Directionality in Surface Point Clouds.} The manifold characteristic of molecular surfaces introduces additional directional information when considering pairwise or ternary interactions among connected particles. To be specific, for each neighboring point pair $(i, j)$, two intersecting planes (see Fig.~\ref{fig:angle}) are formulated with respective normals $\left(\boldsymbol{n}_i, \boldsymbol{n}_j\right)$. We denote the angles between normals and the connecting directed line of two points $(\boldsymbol{x}_{ij}, \boldsymbol{x}_{ji})$ by $\varphi_{\mathbf{n}_i ij}=\angle \boldsymbol{n}_i \boldsymbol{x}_{ij}$ and $\varphi_{\mathbf{n}_j ji}=\angle \boldsymbol{n}_{j} \boldsymbol{x}_{ji}$. We denote the dihedral angle between two half-phases as $\theta_{\mathbf{n}_i ij \mathbf{n}_j}=\angle  \mathbf{n}_i \mathbf{n}_j \perp \boldsymbol{x}_{ij}$. In addition to the common distance $\|\boldsymbol{x}_{ij}\|^2$, these three angles provide a more comprehensive view of understanding the relative position of $(q_i, q_j)$ lying in the surface manifold $Q$, which will also be incorporated into our surface modeling. For instance, for different values of $\left(\varphi_{\mathbf{n}_i ij},\varphi_{\mathbf{n}_j ji}, \theta_{\mathbf{n}_i ij \mathbf{n}_j}\right)$, a triplet of $(\frac{\pi}{2},\frac{\pi}{2},0)$ indicates a perfectly smooth region, while a triplet of $(\pi,\pi,\pi)$ implies a severely sharp and steep curve. 

\paragraph{Equivariant Surface Encoder.} Finally, we draw inspiration from prevalent and modern equivariant algorithms~\citep{gasteiger2020directional,gasteiger2020fast,satorras2021n,zhang2023equipocket} and propose a \textbf{s}urface-based \textbf{e}quivariant \textbf{m}essage \textbf{p}assing (SEMP) as the encoder of $\mathcal{F}_{\boldsymbol{\vartheta}}(\cdot)$. Our SEMP architecture is roto-translation equivariant and leverages both directional and curvature information. 
To begin with, by setting an interaction cutoff $c_{\text {int}}$, we calculate the 3D spherical Fourier-Bessel bases $\left(\boldsymbol{a}_{\mathrm{SBF}}^{(\mathbf{n}_i ij)}, \boldsymbol{a}_{\mathrm{SBF}}^{(\mathbf{n}_j ji)}\right)\in 2\times \mathbb{R}^{N_{\text{CBF}}\times N_{\text{SBF}}\times N_{\text{RBF}}}$ for two angles $\varphi\in \left[\varphi_{\mathbf{n}_i ij}^{(l)}, \varphi_{\mathbf{n}_j ji}^{(l)}\right]$ to integrate orientation knowledge between each interactive particles in the surface:
\begin{equation}
\begin{split}
    a_{\mathrm{SBF}, ovt}^{(l)}&\left(\boldsymbol{x}_{ij}^{(l)}, \varphi, \theta_{\mathbf{n}_i ij\mathbf{n}_j}^{(l)}\right) =\\
    &\sqrt{\frac{2}{c_{\mathrm{int}}^{3} j_{o+1}^{2}\left(z_{ov}\right)}}
    j_o\left(\frac{z_{ov}}{c_{\mathrm{int}}} \left\|\boldsymbol{x}_{ij}^{(l)}\right\|^2 \right) Y_o^t\left(\varphi_{\mathbf{n}_i ij}^{(l)}, \theta_{\mathbf{n}_i ij\mathbf{n}_j}^{(l)} \right),
    \end{split}
\label{equ:sbf}    
\end{equation} 
where $o\in [N_{\text{CBF}}]$, $v\in [N_{\text{SBF}}]$, and $t\in [N_{\text{RBF}}]$ control the degree, root, and order of the radial basis functions, respectively. $\|\mathbf{x}_{ij}\|$ denotes the Euclidean distance between surface points $i$ and $j$. Besides, $j_{o}(\cdot)$ is the $o$-th degree spherical Bessel functions and $z_{ov}$ is its corresponding $v$-th root. $Y_{o}^t(\cdot)$ is the real $o$-th degree and $t$-th order spherical harmonics. 
Equ.~\ref{equ:sbf} can be boiled down to a joint 2D basis if the order $t$ is set to 0. By using $Y_{o}^0(\cdot)$, we obtain the 2D representation $\boldsymbol{a}_{\text{SBF}}^{(\mathbf{n}_i ij \mathbf{n}_j)}\in \mathbb{R}^{N_{\text{CBF}}\times N_{\text{SBF}}}$ based on $\theta_{\mathbf{n}_i ij\mathbf{n}_j}^{(l)}$.  

Remarkably, those 2D/3D spherical Fourier-Bessel representations $\boldsymbol{a}_{\mathrm{SBF}}^{(\mathbf{n}_i ij)}$, $\boldsymbol{a}_{\mathrm{SBF}}^{(\mathbf{n}_j ji)}$, and $\boldsymbol{a}_{\text{SBF}}^{(\mathbf{n}_i ij \mathbf{n}_j)}$ enjoy the roto-translation invariant property due to their exploitation of the relative distance as well as the invariant angles. Then those directional vectors, along with pointwise curvature, are fed into SEMP to attain the initial messages $\boldsymbol{m}_{ij}$ as:  
\small
\begin{equation}
    \boldsymbol{m}_{ij} = f_m\left(\boldsymbol{h}_i^{(l)}, \boldsymbol{h}_j^{(l)}, \boldsymbol{\psi}_i^{(l)}, \boldsymbol{\psi}_j^{(l)}, \boldsymbol{e}_{\text{RBF}}^{(ij)}, \boldsymbol{a}_{\text{SBF}}^{(\mathbf{n}_i ij)}, \boldsymbol{a}_{\text{SBF}}^{(\mathbf{n}_j ji)}, \boldsymbol{a}_{\text{SBF}}^{(\mathbf{n}_i ij \mathbf{n}_j)} \right),
\end{equation}
\normalsize
where $f_m(\cdot)$ is a multi-layer perceptron (MLP) appended with an activation function like SiLU~\citep{nwankpa2018activation}. $\boldsymbol{e}^{(ij)}_{\text{RBF}}$ is the radial basis function representation of the interatomic distance $\| \boldsymbol{x}_{ij}\|^2$. Then a softmax is employed to reweight the messages:
\small
\begin{equation}
    \boldsymbol{m}_{ij}' = a_{ij} \cdot \boldsymbol{m}_{ij},\quad a_{ij}=\frac{\exp(\boldsymbol{W}_{\mathrm{m}}\boldsymbol{m}_{ij} + \boldsymbol{b}_{\mathrm{m}})}{\sum_{k\in\mathcal{N}_{(i)}}\exp(\boldsymbol{W}_{\mathrm{m}}\boldsymbol{m}_{ik} + \boldsymbol{b}_{\mathrm{m}})}
\end{equation}
\normalsize
where the weight matrix $\boldsymbol{W}_{\mathrm{m}}\in\mathbb{R}^{\phi_{m}\times 1}$ and vector $\boldsymbol{b}_{\mathrm{m}}\in\mathbb{R}$ are learnable. After that, messages are propagated from the vicinity of each point $q_i$ to update its node feature as well as coordinates: 
\begin{align}
    \boldsymbol{h}_{i}^{(l+1)} &= f_{h}\left(\boldsymbol{h}_{i}^{(l)}, \boldsymbol{\psi}_i^{(l)}, \sum_{j\in \mathcal{N}_{(i)}} \boldsymbol{m}_{ij}'\right), \\
    \quad \boldsymbol{x}_{i}^{(l+1)} &=\boldsymbol{x}_{i}^{(l)}+\frac{1}{\|\mathcal{N}_{(i)}\|} \sum_{j\in \mathcal{N}_{(i)}} f_{x}\left(\boldsymbol{m}_{ij} \right) \boldsymbol{x}_{ij}^{(l)},   
\end{align}
where $f_{h}(\cdot)$ is another MLP and $f_x:\mathbb{R}^{\phi_m}\rightarrow \mathbb{R}$ transforms $\boldsymbol{m}_{ij}$ into a scalar score to control the impact of directional vector $\boldsymbol{x}_{ij}^{(l)}$. Notably, as the position of each point $\boldsymbol{x}_i^{(l)}$ is moving as the layer $l\in [L]$ goes deeper with $\boldsymbol{x}_i^{(0)}=\boldsymbol{x}_i$, it is optional but recommended to adjust and recalculate the curvature $\boldsymbol{\psi}_i$ and relevant angles $\left(\varphi_{\mathbf{n}_i ij}, \varphi_{\mathbf{n}_j ji}, \theta_{\mathbf{n}_i ij \mathbf{n}_j}\right)$ simultaneously. As angles $\left(\varphi_{\mathbf{n}_i ij}, \varphi_{\mathbf{n}_j ji}, \theta_{\mathbf{n}_i ij \mathbf{n}_j}\right)$ depend on each normal vector pair ($\boldsymbol{n}_i$ and $\boldsymbol{n}_j$), we adopt the local least fitting method~\citep{mitra2003estimating} to estimate and renew $\{\boldsymbol{n}_i\}_{i=1}^m$. In specific, for $q_i$'s updated coordinates $\boldsymbol{x}^{(l)}_i$ at the $l$-th layer, we compute the covariance $\boldsymbol{\Sigma}^{(l)}$ according to Equ.~\ref{equ:covariance} and decompose it to obtain three sorted eigenvalues as well as their corresponding eigenvectors $(\boldsymbol{\nu}_1, \boldsymbol{\nu}_2, \boldsymbol{\nu}_3)$. Then $\boldsymbol{\nu}_3$ with the least eigenvalue is selected as the normal vector $\boldsymbol{n}^{(l)}$ at the $l$-th layer. 

All geometric quantities used as inputs to the message function, including radial distances $r_{ij}$, curvature descriptors $\boldsymbol{\psi}_i$, and angular features $(\varphi_{\mathbf{n}_i ij},\varphi_{\mathbf{n}_j ji}, \theta_{\mathbf{n}_i ij \mathbf{n}_j})$, are invariant under global rigid transformations. Equivariance is preserved by propagating directional information exclusively through relative displacement vectors $\mathbf{x}_{ij}$ in the coordinate update, ensuring SE(3)-equivariant behavior by construction.


\subsection{Reprogramming Protein Language Models}
\paragraph{PEFT for SurfDesign.} Recent works have explored the possibility of transforming PLMs~\citep{rives2021biological,lin2022language,hu2022exploring} into protein design models, and massive evidence demonstrates that the emergent evolutionary knowledge hidden in those PLMs can vastly facilitate the structure-conditioned protein design. Concretely, LM-Design~\citep{zheng2023structure}, InstructPLM~\citep{qiu2024instructplm}, KW-Design~\citep{gao2023kw}, and VFN-IF-ESM~\citep{mao2023modeling} report improvements in CATH 4.2 of 10.8\% (AAR $50.22\% \rightarrow 55.65\%$), 73.9\% (perplexity $10.28 \rightarrow 2.68$), 14.4\% (AAR $54.74\% \rightarrow 62.67\%$), and 17.6\% (AAR $51.66\% \rightarrow 60.77\%$), respectively. 

Motivated by this progress, we also use PLMs as the decoder for $\mathcal{F}_{\boldsymbol{\vartheta}}(\cdot)$ and stack several parameter-efficient fine-tuning (PEFT) techniques to fully exploit the potential of PLMs and significantly reduce the memory footprint. 
Specifically, we utilize a hybrid PEFT method combined with a structural adapter and LoRA~\citep{hu2021lora} with a rank of $r=4$ and a scaling constant of $\alpha=8$. It is worth noting that there is still no consensus on which type of PEFT strategy is most suitable for PLMs~\citep{sledzieski2024democratizing}, and we have found our hybrid mechanism to be more effective than a single strategy for surface-conditioned protein design. 

\paragraph{Training.} We employ the conditional masked language modeling (CMLM)~\citep{zheng2023structure} to better accommodate PLMs that are tasked with MLM~\citep{devlin2018bert} as the training objective. Given the surface $\mathcal{Q}$, CMLM decomposes the sequence into masked and observed ones as $\mathcal{S} = \mathcal{S}_{\text{masked}}\cup \mathcal{S}_{\text{obs}}$ and assumes a conditional independence over identities of target residues $s_i\in \mathcal{S}_{\text{masked}}$. Then it requires the model to predict a set of target amino acids $\mathcal{S}_{\text{masked}}$ from the remaining observed residues $\mathcal{S}_{\text{obs}}$: 
\small
\begin{equation}
    p(\mathcal{S}_{\text{masked}} | \mathcal{S}_{\text{obs}}, \mathcal{Q}; \theta) = \Pi_{s_i\in\mathcal{S}_{\text{masked}}} p(s_i | \mathcal{S}_{\text{obs}}, \mathcal{Q}; \theta)
\end{equation}
\normalsize
where $\mathcal{S}_{\text{masked}}$ is randomly masked. We leverage a coarse-to-fine manner to reconstruct a protein's native sequence from its corrupted version. We also explore this inference scheme with iterative refinement~\citep{savinov2021step} but discover no benefit.  

\section{Experiments}
SurfDesign unifies surface-conditioned inverse folding and functional protein design within a single geometric framework; inverse folding serves as a stress test of geometric compatibility, while binding and catalysis assess functional utility in the absence of ground-truth sequence data. Unless otherwise specified, \emph{Average} denotes a sample-weighted average over the entire test set rather than a simple mean over targets. More experimental details, dataset statistics, and additional results are elaborated in App.~\ref{app:exp_details}. 

\subsection{Protein Binder Design}

\paragraph{Dataset and Setups.}
We focus on optimizing proteins for high-affinity binding under functional and stability constraints and evaluate our method on the benchmark introduced by~\citet{song2024surfpro}, in which the task is to design protein binders that strongly interact with given target receptors. The dataset comprises experimentally validated positive binder-target pairs from six categories, curated in~\citet{bennett2023improving}. Following prior work, we use AlphaFold2 predicted aligned error (AF2 pAE\_interaction) as the evaluation metric, as it quantitatively reflects interface confidence and effectively separates positive binders from negative ones under a fixed binding geometry, without conflating docking-search effects inherent to end-to-end multimer prediction. A designed binder is considered successful if its AF2 pAE\_interaction is lower than that of the corresponding native positive binder. As a control, we include sampled non-binding sequences of the same length as negative binders. All models, including ours, are fine-tuned on the same binder design dataset derived from CATH~4.2 pretraining to ensure a fair comparison.
\begin{table}[t]
\centering
\caption{AF2 pAE\_interaction ($\downarrow$) for all models in the binder design task. The AF2 pAE\_interaction for randomly sampled negative binders of the same length is also provided, along with the positive ones. }
\resizebox{\columnwidth}{!}{%
\begin{tabular}{llllllll} \toprule
    Models & InsulinR & PDGFR & TGFb & H3 & IL7Ra & TrkA & Average \\ \midrule 
    Positive Binder & $\mathbf{5.9996}$ & $\mathbf{14.1366}$ & $\mathbf{15.4884}$ & $\mathbf{21.2631}$& $\mathbf{20.9102}$ & $\mathbf{10.2791}$ & $\mathbf{14.7061}$ \\
    Negative Binder & 19.7167 & 18.0937 & 23.2664 & 22.4556 & 26.0540 & 24.7567 & 21.1335 \\ \midrule 
    Random Baselines & 19.9880 & 21.2690 & 21.4971 & 24.4997 & 24.1541 & 23.1147 & 22.2020 \\
    ProteinMPNN & 18.3393 & 25.2919 & 25.8559 & 24.5968 & 25.5278 & 27.0980 & 23.4462 \\
    PiFold & 12.9809 & 21.8230 & 24.4737 & 23.3924 & 26.6738 & 19.7172 & 20.5785 \\
    LM-DESIGN & 13.6440 & 22.0749 & 23.3725 & 23.8332 & 24.3937 & 22.3987 & 20.7728 \\
    SurfPro & 10.2608 & 17.9862 & 17.7364 & 21.2916 & 20.8594 & 10.6535 & 16.9485 \\ 
    SurfPro-Pretrain &  11.2530 & 18.4141 & 15.4011 & 22.2704 & 20.5700 & 21.3515 & 17.6699 \\ \midrule 
    \grow{SurfDesign} & $\mathbf{8.9827}$ & $\mathbf{16.3462}$ & $\mathbf{17.4338}$ & $\mathbf{21.0642}$ & $\mathbf{20.8207}$ & $\mathbf{10.4288}$ & $\mathbf{15.8460}$ \\ \bottomrule
\end{tabular}}
\label{tab:binder_design}
\end{table}

\begin{table}[t]
\centering
\caption{Success rate (\%, $\uparrow$) of different models on the binder design task.
SurfDesign achieves the highest overall success rate.}
\resizebox{\columnwidth}{!}{%
\begin{tabular}{lccccccc}\toprule
\multirow{2}{*}{Model}  & \multicolumn{3}{c}{Seen Class} & \multicolumn{3}{c}{Zero-Shot} & \multirow{2}{*}{Average} \\ \cmidrule(lr){2-4} \cmidrule(lr){5-7}
& InsulinR & PDGFR & TGFb & H3 & IL7Ra & TrkA & \\ \midrule
ProteinMPNN        &  3.22 &  5.71 & 20.71 & 18.68 & 24.10 &  7.50 & 11.96 \\
PiFold             & 20.64 &  3.57 & 19.19 & 29.21 & 22.85 & 20.00 & 19.32 \\
LM-DESIGN          &  7.74 & 15.00 & 15.71 & 22.29 & 24.28 & 25.00 & 16.37 \\
SurfPro            & 31.57 & 19.99 & 11.61 & 23.21 & 19.28 & 25.00 & 22.29 \\
SurfPro-Pretrain   &  5.48 & 27.14 & 33.57 & 37.63 & 38.57 & 25.00 & {26.22} \\ \midrule 
 \grow{SurfDesign} & 34.87 & 24.28 & 29.46 & 32.89 & 34.28 & 27.50 & \textbf{30.14} \\ \bottomrule
\end{tabular}}
\vspace{-1em}
\label{tab:binder_success}
\end{table}

\paragraph{Results.} As shown in Tab.~\ref{tab:binder_design}, SurfDesign achieves an average AF2 pAE\_interaction of 15.85, the lowest among all models. This represents a clear improvement over the prior surface-conditioned method, SurfPro (16.95), and significantly outperforms non-surface models such as LM-DESIGN (20.77), PiFold (20.58), and ProteinMPNN (23.45). SurfDesign consistently yields top performance on five of six targets: PDGFR (16.35), TGFb (17.43), H3 (21.06), IL7Ra (20.82), and TrkA (10.43). Lower AF2 pAE\_interaction directly reflects increased structural certainty at the designed interface, though it is not a direct measure of binding affinity, indicating that SurfDesign produces binders whose surfaces more consistently support stable receptor engagement. While the improvement over SurfPro is moderate, it reflects the benefits of enhanced surface modeling and improved geometric representation of interaction sites. In contrast, negative and random binders yield much higher AF2 pAE\_interaction scores (21.13 and 22.20, respectively), further validating the discriminative power of this metric. These findings support the hypothesis that surface-informed representations can enhance the design of functional proteins with superior binding fidelity. SurfDesign thus represents a promising direction for high-accuracy functional protein generation grounded in surface geometry.

\subsection{Enzyme Design}
Enzyme design poses a stricter test than binder design, as successful generation requires precise pocket geometry and localized physicochemical patterns rather than global shape complementarity.

\paragraph{Dataset and Setups.}   We evaluate SurfDesign on the \emph{enzyme design} task, where the objective is to generate enzyme sequences that bind specific small-molecule substrates. It tests whether surface-conditioned generation can capture fine-grained geometric and physicochemical patterns required for enzyme--substrate interactions. We adopt the benchmark compiled by~\citet {kroll2023general} comprising five enzyme categories, each associated with a distinct substrate. To prevent data leakage, we explicitly exclude all enzymes that appear in the CATH~4.2 dataset. For enzyme categories containing more than 100 samples, we perform clustering-based splitting and randomly partition the data into training, validation, and test sets using an $8{:}1{:}1$ ratio. For categories with fewer samples, all enzymes are used exclusively for testing in a zero-shot setting. 

To assess enzyme-substrate binding affinity, we adopt the Enzyme-Substrate Potential (ESP) score proposed by~\citet{kroll2023general}. The ESP model predicts enzyme-substrate interactions with approximately 91\% accuracy across multiple benchmarks, with higher values indicating greater predicted enzyme-substrate compatibility. We use the official implementation released by~\citet{kroll2023general} to compute ESP scores for all designed enzymes. Following the protocol used in protein binder design, we report (i) the average ESP score obtained via greedy decoding, and (ii) the average success rate computed from sequences generated by sampling with temperature $0.1$. For each enzyme-substrate pair, the success rate is defined as the fraction of generated sequences whose ESP score is better than that of the corresponding native enzyme.

Consistent with the binder design setup, we fine-tune all baseline models on the enzyme design dataset, starting from models pretrained on the inverse folding task. We additionally report results for a random baseline and SurfPro-Pretrain under the same settings as the binder design. Importantly, the pretraining corpus for SurfPro-Pretrain explicitly excludes all enzymes used in this evaluation to avoid any potential data leakage.

\paragraph{Results.} Quantitatively, SurfDesign improves over prior surface and non-surface baselines on both metrics. For ESP score (Tab.~\ref{tab:enzyme_esp}), SurfDesign achieves the best overall average of \textbf{0.9058}, compared to SurfPro (\textbf{0.8931}) and ProteinMPNN (\textbf{0.8676}), and remains competitive with LM-DESIGN (\textbf{0.9037}) despite LM-DESIGN benefiting from large-scale PLM pretraining. For success rate (Tab.~\ref{tab:enzyme_success}), SurfDesign attains the highest overall success rate among the compared methods at \textbf{47.30\%}, exceeding SurfPro (\textbf{42.23\%}) and SurfPro-Pretrain (\textbf{43.63\%}), and outperforming PiFold (\textbf{40.65\%}) and LM-DESIGN (\textbf{37.58\%}). Notably, gains persist in the zero-shot setting (C00001), where SurfDesign achieves \textbf{34.36\%} success, compared with SurfPro's \textbf{33.55\%}. These suggest that more expressive manifold-aware surface representations can better capture localized pocket geometry and physicochemical patterns critical for enzyme-substrate interactions.

\begin{table}[t]
\centering
\caption{ESP score ($\uparrow$) of different models under greedy decoding on the enzyme design task. Substrates are denoted by their KEGG IDs. }
\resizebox{1\columnwidth}{!}{%
\begin{tabular}{lccccc c} \toprule
\multirow{2}{*}{Model}  & \multicolumn{4}{c}{Seen Class} & Zero-Shot  & \multirow{2}{*}{Average} \\ \cmidrule(lr){2-5} \cmidrule(lr){6-6}
& C00002 & C00677 & C00019 & C00003 & C00001 & \\ \midrule
Real Enzyme          & 0.9573 & 0.8642 & 0.4497 & 0.8076 & 0.9892 & 0.9091 \\\midrule
Random Baseline      & 0.5523 & 0.2475 & 0.1673 & 0.4705 & 0.7891 & 0.5292 \\
ProteinMPNN          & 0.9711 & 0.7375 & 0.2614 & 0.6699 & 0.9763 & 0.8676 \\
PiFold               & 0.9142 & 0.8816 & 0.4296 & 0.8212 & 0.9616 & 0.8865 \\
LM-DESIGN            & 0.9498 & 0.8836 & 0.4585 & 0.8078 & 0.9650 & 0.9037 \\
SurfPro              & 0.9264 & 0.8921 & 0.3892 & 0.7631 & 0.9772 & 0.8931 \\
SurfPro-Pretrain     & 0.9376 & 0.8631 & 0.3949 & 0.7668 & 0.9691 & 0.8900\\ \midrule 
 \grow{SurfDesign} & 0.9487 & 0.9012 & 0.4523 & 0.8189 & 0.9801 & \textbf{0.9058}  \\ \bottomrule
\end{tabular}}
\label{tab:enzyme_esp}
\end{table}

\begin{table}[t]
\centering
\caption{Success rate (\%, $\uparrow$) of different models on the enzyme design task. Substrates are denoted by their KEGG database IDs. SurfDesign achieves the highest average success rate.}
\resizebox{1\columnwidth}{!}{%
\begin{tabular}{lccccc c} \toprule
\multirow{2}{*}{Model} & \multicolumn{4}{c}{Seen Class} & Zero-Shot  & \multirow{2}{*}{Average} \\\cmidrule(lr){2-5} \cmidrule(lr){6-6}
& C00002 & C00677 & C00019 & C00003 & C00001 & \\ \midrule
ProteinMPNN        & 47.54 & 31.63 & 58.82 & 44.72 & 27.65 & 39.23 \\
PiFold             & 48.54 & 41.72 & 58.29 & 37.54 & 24.97 & 40.65 \\
LM-DESIGN          & 45.00 & 42.54 & 43.76 & 53.63 & 20.13 & 37.58 \\
SurfPro            & 43.36 & 46.00 & 59.41 & 45.45 & 33.55 & 42.23 \\
SurfPro-Pretrain   & 50.90 & 41.81 & 52.94 & 36.36 & 34.21 & 43.63 \\ \midrule 
 \grow{SurfDesign} & 52.73 & 45.45  & 60.18 & 51.82 & 34.36 & \textbf{47.30} \\ \bottomrule
\end{tabular}}
\label{tab:enzyme_success}
\end{table}

\subsection{Inverse Folding as Compatibility Analysis}
We include inverse folding results to assess whether surface-conditioned generation preserves global fold compatibility, rather than treating it as a primary design objective. Various benchmarks are used to design fixed-backbone protein sequences, including single-chain monomers and multi-chain protein complexes. 

\paragraph{Baselines and Datasets.} A wide variety of approaches are established for comparison, most of which are open source. Among them, StructGNN~\citep{ingraham2019generative}, GraphTrans~\citep{ingraham2019generative}, GVP~\citep{jing2020learning}, ProteinMPNN~\citep{dauparas2022robust}, AlphaDesign~\citep{gao2022alphadesign}, PiFold~\citep{gao2022pifold}, UniIF~\citep{gao2024uniif}, etc. are GNN-based algorithms, while DenseCPD~\citep{qi2020densecpd} is CNN-based. DPLM~\citep{wang2024diffusion}, InstructPLM~\citep{qiu2024instructplm}, LM-Design, KW-Design and VFN-IF-ESM~\citep{mao2023modeling} leverage and integrate PLMs' knowledge. GRADE-IF~\citep{yi2023graph} and DMRA~\citep{wang2024diffusion} rely on diffusion. SurfPro~\citep{song2024surfpro} is a surface-conditioned framework. Using the same splitting strategy as the compared systems~\citep{dauparas2022robust}, proteins in CATH 4.2 were partitioned into 18,024/608/1,120 samples for training, validation, and testing, respectively. To compare with ESM-IF~\citep{hsu2022learning}, structures in CATH 4.3 were split into 16,153/1,457/1,797 samples for training, validation, and testing, respectively. To provide a head-to-head comparison with ESM-IF, no additional data, such as AF2DB~\citep{varadi2022alphafold}, was used to train SurfDesign. To evaluate generative quality thoroughly, we report perplexity and the median AAR rate in the short-chain, single-chain, and all-chain settings, as usual. 
The multi-chain protein design employs the dataset curated by~\citet{dauparas2022robust}, which was preprocessed by clustering sequences at 30\% sequence identity, yielding 25,361 clusters. Following ProteinMPNN’s setup, the clusters were randomly divided into 23,358/1,464/1,539 samples for training, validation, and testing, respectively. This strategy ensures that none of the target chain's chains or biounits were present in the other two sets. 
\begin{table}[t]
\caption{Sequence design performance and ablation studies on CATH 4.2 held-out test split. The \textbf{best performance} is shown in bold, while the \underline{best baseline} is indicated with an underline. ESM-IF is tested on CATH 4.2, although it was originally trained and evaluated on CATH 4.3.}
\label{tab:cath2}
\centering
\resizebox{1\columnwidth}{!}{%
\begin{tabular}{lc|ccc|ccc}\toprule
\multirow{2}{*}{ \textbf{Models} } & \textbf{Trainable/Total} & \multicolumn{3}{c|}{ \textbf{Perplexity} ($\downarrow$)} & \multicolumn{3}{c}{\textbf{Median AAR} ($\uparrow$) } \\
& Params. & Short & Single-chain & All & Short & Single-chain & All \\ \midrule
StructGNN~\citep{ingraham2019generative} &  1.4M / 1.4M & 8.29 & 8.74 & 6.40 & 29.44 & 28.26 & 35.91 \\
GraphTrans~\citep{ingraham2019generative} &  1.56M / 1.56M & 8.39 & 8.83 & 6.63 & 28.14 & 28.46 & 35.82 \\
GCA~\citep{tan2023global} &  2.1M / 2.1M & 7.09 & 7.49 & 6.05 & 32.62 & 31.10 & 37.64 \\
GVP~\citep{jing2020learning} &  1.0M / 1.0M & 7.23 & 7.84 & 5.36 & 30.60 & 28.95 & 39.47 \\
AlphaDesign~\citep{gao2022alphadesign} &  3.6M / 3.6M & 7.32 & 7.63 & 6.30 & 34.16 & 32.66 & 41.31 \\
ProteinMPNN~\citep{dauparas2022robust} &  1.9M / 1.9M & 6.21 & 6.68 & 4.61 & 36.35 & 34.43 & 45.96 \\
ESM-IF~\citep{hsu2022learning}&  142M / 142M & 6.93 & 6.65 & 3.96 & 35.28 & 33.78 & 48.95 \\
PiFold ~\citep{gao2022pifold}&  6.6M / 6.6M & 6.04 & 6.31 & 4.55 & 39.84 & 38.53 & 51.66 \\
LM-Design-MPNN~\citep{zheng2023structure} &  5.0M / 659M & 7.01 & 6.58 & 4.41 & 35.19 & 40.00 & 54.41 \\
LM-Design-PiFold~\citep{zheng2023structure}  &  11.9M / 664M & 6.77 & 6.46 & 4.52 & 37.88 & 42.47 & 55.65 \\
DPLM~\citep{wang2024diffusion} &  5.0M / 659M & -- & -- & -- & -- & -- & 54.54 \\ 
InstructPLM~\citep{qiu2024instructplm} &  89.1M / 6.6B & $\underline{3.22}$ & ${3.17}$ & $\underline{2.68}$ & $\underline{61.59}$ & $\underline{59.29}$ & 57.51 \\
KW-Design~\citep{gao2023kw} &  6.4M / 798M & 5.48 & 5.16 & 3.46 & 44.66 & 45.45 & 60.77 \\ 
VFN-IF~\citep{mao2023modeling} & 5.4M / 5.4M & 5.70 & 5.86 & 4.17 & 41.34 & 40.98 & 54.74 \\ 
VFN-IF-ESM~\citep{mao2023modeling} &  5.4M / 15B & 4.92 & 4.22 & 3.36 & 50.00 & 52.13 & 62.67 \\
SurfPro~\citep{song2024surfpro} & 5.8M / 5.8M & -- & -- & 3.13 & -- & -- & 57.78 \\  
PRISM~\citep{mahbub2025prism} &  -- / -- &3.74 & $\underline{2.68}$ & 2.71 & 60.89 &40.98 &60.43 \\ 
GRADE-IF~\citep{yi2023graph} &  -- / -- & 5.49  & 6.21 & 4.35 & 45.27&42.77 & 52.21 \\ 
DMRA~\citep{wang2024diffusion} &  -- / --  & 4.06 & 4.76 & 2.93 & 53.57 & 48.95 & \underline{64.07} \\ 
MapDiff~\citep{bai2025mask} & 14.7M / 14.7M & 3.96&  4.41 &3.43 &54.04 & 49.34& 60.93 \\\midrule
\grow{SurfDesign-backbone} & 5.3M / 656M &  3.28 & 3.11 & 3.16 & 63.45 & 64.32 & 65.12 \\ \midrule 
\grow{SurfDesign (w/o PLMs)} & 5.3M / 5.3M & 3.21 & 3.10 & 3.08 & 62.70 & 64.88 & 65.35 \\ 
\grow{SurfDesign (w/o SEMP)} & 4.8M / 655M & 3.08 & 2.93 & 2.76 & 65.43 & 67.06 & 66.27 \\ 
\grow{SurfDesign} & 5.3M / 656M & $\mathbf{2.43}$ & $\mathbf{2.44}$ & $\mathbf{2.41}$ &$\mathbf{73.74}$ & $\mathbf{75.17}$ & $\mathbf{74.13}$\\  \bottomrule
\end{tabular}}
\end{table}

\paragraph{Single-chain Protein Design.} \label{sec:single_chain}
Tab.~\ref{tab:cath2} and~\ref{tab:cath3} document the results on the CATH~\citep{orengo1997cath} benchmark, where SurfDesign consistently achieves state-of-the-art performance in distinct settings.  Under a controlled CATH-only training setting, SurfDesign is the first surface-conditioned model to exceed 70\% AAR on CATH 4.2 and CATH 4.3, demonstrating its superior ability to restore effective protein sequences. On the full CATH 4.2, SurfDesign achieves a perplexity of 2.41 and an AAR of 74.13\%, outperforming the previous state-of-the-art VFN-IF-ESM by 28.27\% and 18.28\%, respectively. It also induces AAR improvements of 19.72\% and 26.78\% on the short and single-chain subsets, respectively. Furthermore, SurfDesign surpasses SurfPro, another surface-conditioned algorithm, by 23.00\% and 28.29\% in the overall metrics, respectively. The outstanding phenomenon also exists for the CATH 4.3 benchmark, where SurfDesign outperforms the strongest competitor, KW-Design, by 10.60\% in perplexity and 19.49\% in AAR. To summarize, SurfDesign enhances surface-conditioned sequence generation with greater efficiency, thanks to the significant advancements and open-source contributions from the entire community, building on the foundation laid by previous pioneers. To address data leakage concerns and demonstrate that SurfDesign's >70\% recovery results more from manifold-aware geometric reasoning than from privileged sequence information, our backbone-only variant achieves 65.12\% AAR and still outperforms backbone-only baselines, while testing on noisy ESMFold-predicted structures maintains 68.5\% AAR, demonstrating robustness independent of native side-chain accuracy.

\paragraph{Multi-chain Protein Complex Design.} \label{sec:multi_chain}
A protein functions only when it docks, associates, and interacts with other macromolecules, forming multi-chain protein complexes. Thus, studying protein sequence design for multi-chain assembled structures is crucial, motivating us to assess whether SurfDesign can design a protein complex. From Appendix Tab.~\ref{tab:multi-chain}, we conclude that the AAR is generally higher for longer proteins, and all models achieve higher AAR rates on PDB than CATH datasets. More importantly, SurfDesign achieves the best performance, with an AAR exceeding 80\%. This phenomenon indicates that SurfDesign can design both single-chain proteins and multi-chain complexes. This makes SurfDesign more versatile in the categories and scenarios in which it can be deployed, creating opportunities to use it to design specific protein complexes.

\paragraph{Zero-shot Generalization to New Protein Families.}
TS50 and TS500~\citet{li2014direct} are commonly used independent test sets to assess model generalization on unseen proteins. Towards this goal, we evaluate SurfDesign trained on CATH 4.2 and 4.3, respectively, and report the results in Tab.~\ref{tab:ts50_500}. We find that SurfDesign outperforms prior studies by a large margin across all benchmarks. Specifically, it achieves a perplexity of $2.05$ and an AAR rate of $82.16$ on TS50, outperforming the previous state-of-the-art algorithm, VFN-IF-ESM, by 18.65\% and 12.08\%, respectively. Meanwhile, on the TS500 dataset, SurfDesign obtains a perplexity of $1.98$ and an AAR rate of $84.70$. These numbers are better than VFN-IF-ESM by 22.04\% and 16.80\%, respectively. In addition, for those trained in CATH 4.3, SurfDesign consistently achieves the best. In a nutshell, SurfDesign is the first to transcend 82\% and 84\% AAR on the TS50 and TS500. 
\begin{table}[t]
    \centering
    \vspace{-1em}
    \caption{Structure recovery based on the self-consistent protocol~\citep{yim2023se}. $\ddagger$: results are quoted from~\citet{mao2023modeling}.}
    \vspace{-1em}
    \label{tab:structure} 
    \resizebox{0.85\columnwidth}{!}{
    \begin{tabular}{l|cccc} \toprule
        \textbf{Metrics} & PiFold$^\ddagger$ & LM-Design$^\ddagger$ & VFN-IF-ESM$^\ddagger$ & \cellcolor{blue!25}SurfDesign \\ \midrule
        \textbf{scTM $> 0.5$} &  90.98\% & 89.42\% & 93.29 \%  & \cellcolor{blue!25}\textbf{96.17\%}\\ 
        \textbf{scRMSD $< 2.0$} & 60.35 \% & 58.41\% & 64.16\% & \cellcolor{blue!25}\textbf{72.83\%} \\ \bottomrule
    \end{tabular}}
    \vspace{-1em}
\end{table}

\subsection{Discussion and Analysis}
By elevating molecular surfaces to first-class conditioning signals, SurfDesign shifts protein design from a fold-centric to an interaction-centric paradigm. This perspective aligns naturally with emerging pipelines that generate or refine structures before sequence design, and suggests a modular future in which surface geometry, sequence priors, and functional objectives can be composed rather than entangled.

\paragraph{Ablation Studies.} We systematically investigate the contributions of SurfDesign's components, shown in Tab.~\ref{tab:cath2}. It can be observed that the knowledge of PLMs provides a large improvement of 13.43\% in AAR ($65.35\% \rightarrow 74.13\%$) and a decrease of 24.29\% in perplexity ($3.21 \rightarrow 2.43$). Moreover, the incorporation of directionality and curvatures also contributes to the superiority of SurfDesign, with improvements of 11.86\% in AAR and 12.68\% in perplexity. 


\begin{wraptable}{r}{0.55\columnwidth}
    \vspace{-1.5em}
    \centering
    \caption{Evaluation on the surface recovery on CATH 4.2. }
    \label{tab:surf} 
    \resizebox{0.55\columnwidth}{!}{
    \begin{tabular}{l|ccc} \toprule
        \textbf{Models} & \textbf{IoU} ($\uparrow$) & \textbf{CD} ($\downarrow$) & \textbf{NC} ($\uparrow$) \\ \midrule
        LM-Design &  0.90 & 5.972 & 0.4236 \\ 
        VFN-IF-ESM & 0.92 & 4.688  & 0.4859 \\  \midrule
        \grow{SurfDesign} & \textbf{0.98} & \textbf{2.873} & \textbf{0.6241}\\ \bottomrule
    \end{tabular}}
    \vspace{-1em}
\end{wraptable}
\paragraph{Surface Recovery.} The ultimate goal of our surface-conditioned design is to generate proteins with higher surface similarity of key regions, such as the binding or interaction site~\citep{lai2024interformer}. To measure the similarity between two 3D molecular shapes, we use three evaluation metrics~\citep{sun2024dsr} commonly used in 3D modeling from three aspects: volume, distance, and normal vectors. They are Volumetric Intersection over Union (IoU), Chamfer distance (CD), and Normal Consistency (NC) (computational details are in App.~\ref{app:surface_compare}). As shown in Tab.~\ref{tab:surf}, SurfDesign reconstructs molecular surfaces well, aligning with the motivation for our surface-conditioned design. Visualization of the generated and ground-truth surfaces is provided in App.~\ref{app:surf_visual}.

\paragraph{Structure Recovery.} We compare SurfDesign with strong baselines in terms of protein structure recovery on CATH 4.2, reported in Tab.~\ref{tab:structure}. Following standard evaluation procedures~\citep{yim2023se,mao2023modeling}, ESMFold was used to predict structures of designed sequences. A case study of visualization comparison using Alphafold-3 is displayed in App.~\ref{app:visual}. Two self-consistent metrics, scTM ($\uparrow$) and scRMSD ($\downarrow$) are leveraged to assess the similarity between desired and designed protein structures. SurfDesign is more likely to generate protein sequences with the expected structures. More analysis of refoldability is elaborated in App.~\ref{app:refoldability}.

\begin{figure}[t]
  \begin{center}
    \includegraphics[width=0.7\columnwidth]{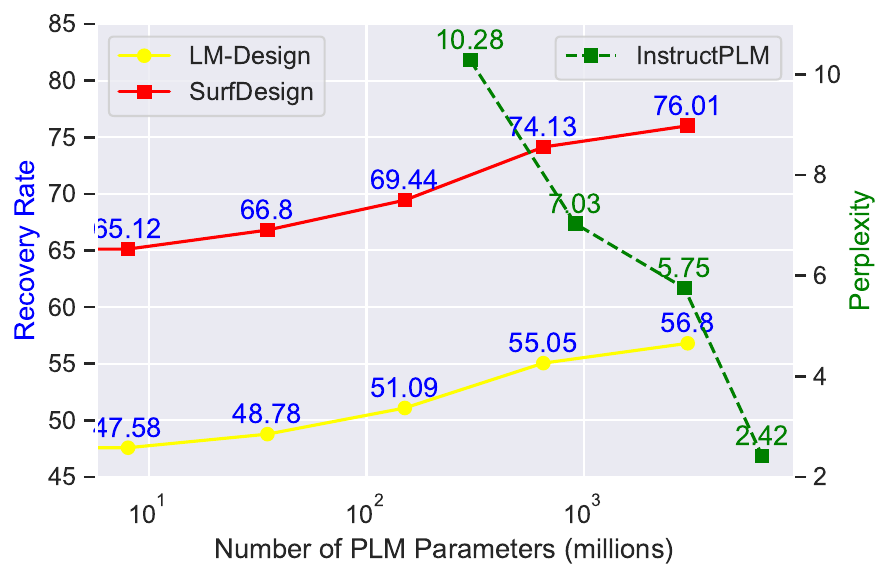}
  \end{center}
  \vspace{-1em}
  \caption{Performance of different PLM scales.} 
  \label{fig:plm}
  \vspace{-1em}
\end{figure}
\paragraph{Scalability of PLMs.} The scaling law w.r.t model sizes of PLMs has recently been studied~\citep{qiu2024instructplm}. To understand the influence of PLM model size on SurfDesign's capacity, we increase the ESM-2 parameter count from 8M to 3B. A similar phenomenon has been discovered in Fig.~\ref{fig:plm}, where the performance of SurfDesign improves as PLMs scale. When integrating knowledge from the largest PLM (3B), SurfDesign achieves a recovery rate of 76.01\% on CATH 4.2. This coincidence highlights the significant potential of integrating surface-conditioned design with state-of-the-art PLMs. 

\begin{wrapfigure}{r}{0.6\columnwidth}
  \vspace{-1.8em}
  \begin{center}
    \includegraphics[width=0.6\columnwidth]{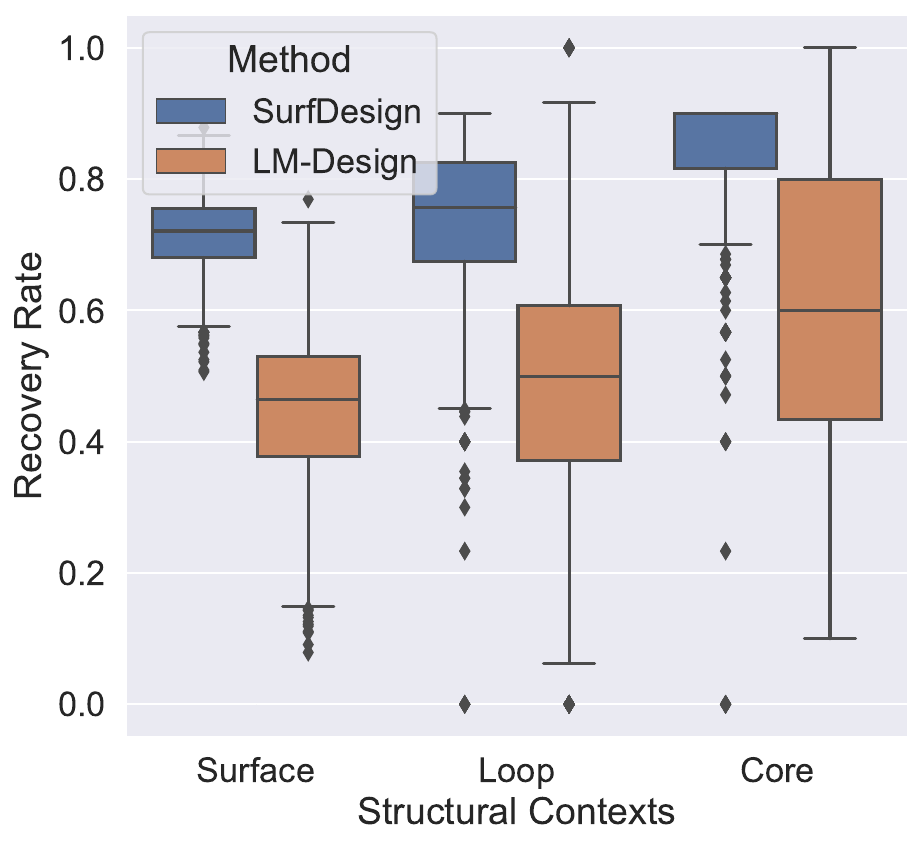}
  \end{center}
  \vspace{-1em}
  \caption{Sequence recovery w.r.t. structural contexts regarding SASA and interaction interface, on CATH 4.2 single-chain proteins.} 
  \label{fig:box}
  \vspace{-4em}
\end{wrapfigure}
\paragraph{Structural Contexts.} We dissect the action mechanism of SurfDesign according to different structural contexts in Fig.~\ref{fig:box}.  Structure-based LM-Design shows high AAR on structurally constrained residues in the folding core, while low AAR in structurally less constrained residues on surface areas and loops. SurfDesign significantly enhances the recovery of structurally constrained and less-constrained residues, particularly those on the surface regions. 

\section{Related Work}
\label{app:related_work}
\paragraph{Structure-based Protein Design.} Advances in AI-driven structure prediction, notably AlphaFold~\citep{jumper2021highly}, have reinvigorated the complementary task of \emph{inverse folding}. Early inverse folding methods formulated the problem as per-residue classification using multi-layer perceptrons (MLPs) with handcrafted structural features. SPIN~\citep{li20143d} combined torsion angles, sequence profiles, and energy descriptors to achieve 30\% AAR on TS50, while SPIN2~\citep{o2018spin2} incorporated backbone angles, contact numbers, and inter-residue distances, improving AAR to 34\%.~\citet{wang2018computational} leveraged backbone dihedrals, solvent-accessible surface area, secondary-structure annotations, and unit direction vectors, attaining 33\% AAR. Subsequent work replaced MLPs with convolutional architectures to better capture spatial context. SPROF~\citep{chen2019improve} applied 2D CNNs to C$\alpha$–C$\alpha$ distance maps, achieving 40.25\% AAR on TS500. In three dimensions, ProDCoNN~\citep{zhang2022ontoprotein} employed a multi-scale 3D CNN to reach 42.2\%, while DenseCPD~\citep{qi2020densecpd} further improved performance to 55.53\%.

Recognizing proteins as inherently graph-structured objects, recent methods adopt GNNs to better respect geometric constraints. GraphTrans~\citep{ingraham2019generative} introduced a graph-attention encoder with an autoregressive decoder, while GVP incorporated geometric vector perceptrons to jointly process scalar and vector features. Building on this foundation, GCA~\citep{tan2023global} integrates global attention across residue graphs; AlphaDesign~\citep{gao2022alphadesign} proposes a streamlined GVP-based encoder with a constraint-aware decoder; ProteinMPNN combines autoregressive decoding with iterative message passing; and PiFold introduces virtual atoms and explicit backbone dihedral modeling. VFN employs learnable vector operations over frame-anchored virtual atoms, pushing AAR to 62.67\%.

Despite steady improvements, limited structural data constrain sequence diversity. To mitigate this, ESM-IF leverages large-scale AlphaFold2 predictions to pretrain a GVP-based model. LM-Design fine-tunes ESM-2 conditioning on pretrained structural embeddings, and InstructPLM incorporates explicit \emph{structure prompts} via cross-modal alignment in ProGen2~\citep{nijkamp2023progen2}. KW-Design enhances low-confidence residues using knowledge from ESM and GearNet~\citep{zhang2022protein}, while recent work~\citep{wang2024diffusion} shows that self-supervised discrete diffusion models can serve as general protein learners for structure-conditioned sequence generation.

\paragraph{Protein Surface Modeling.} 
The characteristics of the molecular surface dictate the type and strength of the interactions that a protein can have with other molecules. It is defined by van der Waals (vdW) radii~\citep{connolly1983analytical} and is commonly represented as meshes derived from signed distance functions. 
MaSIF~\citep{gainza2020deciphering} pioneered the use of mesh-based geometric DL to abstract the internal parts of the protein fold and explore protein interactions. A subsequent study~\citep{sverrisson2021fast} reduced pre-computation costs by modeling molecular surfaces as point clouds, assigning atom categories to each point. Other seminal works have linked protein surfaces with structural information in a multimodal manner~\citep{somnath2021multi}, incorporating comprehensive pretraining strategies~\citep{wu2024surface} using implicit neural representations (INRs)~\citep{park2019deepsdf} for self-supervised learning~\cite{lee2023pre} and dynamic structure modeling~\cite{sun2024dsr}.
Despite these efforts, surface-conditioned design remains underexplored. Recent advancements, such as the work by \citet{gainza2023novo} on expanding MaSIF for \emph{de novo} binder design and SurfPro~\citep{song2024surfpro}, which eliminates the need for handcrafted feature calculations, have begun to address this gap by directly generating functional proteins from surface data.

\paragraph{Parameter-efficient Fine-tuning.} 
Training and storing full copies of large PLMs\citep{lin2022language,rao2019evaluating,elnaggar2020prottrans} for various downstream tasks are increasingly impractical. PEFT techniques~\citep{sledzieski2024democratizing}, such as LoRA~\citep{hu2021lora} and prompt tuning~\citep{lester2021power}, achieve competitive or superior performance compared to full fine-tuning, significantly reducing memory for tasks like interaction prediction and homo-oligomer symmetry prediction.  Recent work integrates structural information into PLMs via PEFT. LM-Design introduces a lightweight adapter to realize structural awareness, referred to as \emph{structural surgery} on PLMs. SES-Adapter~\citep{tan2024simple} integrates structural data by converting it into sequential vectors using tools such as FoldSeek~\citep{van2024fast} and DSSP~\citep{kabsch1983dictionary}, thereby enabling cross-modal attention calculations. It outperforms structure-aware PLMs such as SaProt~\citep{su2023saprot} on standard datasets, including thermostability, metal-ion binding, gene ontology annotations, and subcellular localization prediction.

\section{Limitations and Ethical Considerations}
SurfDesign depends on the quality of input structures and molecular surface reconstructions, and errors in upstream structure prediction may affect performance. While inverse folding and in silico benchmarks assess geometric and functional compatibility, they do not replace experimental validation of binding affinity, stability, or activity. Protein design may pose dual-use risks if misapplied. This work is intended solely for benign scientific and biomedical research. We do not support or evaluate applications related to harmful biological agents and emphasize the importance of responsible use in compliance with institutional biosafety and ethical guidelines.

\section{Conclusion}
In this work, we presented SurfDesign, a surface-conditioned protein design framework that elevates molecular surfaces from auxiliary inputs to first-class conditioning signals for generative protein modeling.  
\section*{Acknowledgments}
This work was supported by the National Natural Science Foundation of China (Grant No. 	62402351).

\bibliographystyle{ACM-Reference-Format}
\bibliography{cite}

@article{trippe2022diffusion,
  title={Diffusion probabilistic modeling of protein backbones in 3d for the motif-scaffolding problem},
  author={Trippe, Brian L and Yim, Jason and Tischer, Doug and Baker, David and Broderick, Tamara and Barzilay, Regina and Jaakkola, Tommi},
  journal={arXiv preprint arXiv:2206.04119},
  year={2022}
}

@article{li2025surffold,
  title={SurfFold: A Unified Model for Protein Inverse Folding by Integrating Surface and Structural Information},
  author={Li, Darong and Shen, Lian and Song, Meijia and Li, Deyi and Liu, Juan and Liu, Xiangrong},
  journal={Bioinformatics},
  pages={btaf666},
  year={2025},
  publisher={Oxford University Press}
}

@article{kroll2023general,
  title={A general model to predict small molecule substrates of enzymes based on machine and deep learning},
  author={Kroll, Alexander and Ranjan, Sahasra and Engqvist, Martin KM and Lercher, Martin J},
  journal={Nature communications},
  volume={14},
  number={1},
  pages={2787},
  year={2023},
  publisher={Nature Publishing Group UK London}
}

@article{yi2023graph,
  title={Graph denoising diffusion for inverse protein folding},
  author={Yi, Kai and Zhou, Bingxin and Shen, Yiqing and Li{\`o}, Pietro and Wang, Yuguang},
  journal={Advances in Neural Information Processing Systems},
  volume={36},
  pages={10238--10257},
  year={2023}
}

@article{xu2025protein,
  title={Protein Inverse Folding From Structure Feedback},
  author={Xu, Junde and Gao, Zijun and Zhou, Xinyi and Hu, Jie and Cheng, Xingyi and Song, Le and Chen, Guangyong and Heng, Pheng-Ann and Qiu, Jiezhong},
  journal={arXiv preprint arXiv:2506.03028},
  year={2025}
}

@article{mahbub2025prism,
  title={PRISM: Enhancing Protein Inverse Folding through Fine-Grained Retrieval on Structure-Sequence Multimodal Representations},
  author={Mahbub, Sazan and Kundu, Souvik and Xing, Eric P},
  journal={arXiv preprint arXiv:2510.11750},
  year={2025}
}

@article{bai2025mask,
  title={Mask-prior-guided denoising diffusion improves inverse protein folding},
  author={Bai, Peizhen and Miljkovi{\'c}, Filip and Liu, Xianyuan and De Maria, Leonardo and Croasdale-Wood, Rebecca and Rackham, Owen and Lu, Haiping},
  journal={Nature Machine Intelligence},
  pages={1--13},
  year={2025},
  publisher={Nature Publishing Group UK London}
}

@inproceedings{li2025joint,
  title={Joint Design of Protein Surface and Backbone Using a Diffusion Bridge Model},
  author={Li, Guanlue and Zhao, Xufeng and Wu, Fang and Laue, S{\"o}ren},
  booktitle={The Thirty-ninth Annual Conference on Neural Information Processing Systems},
  year={2025}
}

@inproceedings{wu2025surface,
  title={Surface-based Molecular Design with Multi-modal Flow Matching},
  author={Wu, Fang and Zhou, Zhengyuan and Jin, Shuting and Zeng, Xiangxiang and Leskovec, Jure and Xu, Jinbo},
  booktitle={Proceedings of the 31st ACM SIGKDD Conference on Knowledge Discovery and Data Mining V. 2},
  pages={3192--3203},
  year={2025}
}

@article{ektefaie2024reinforcement,
  title={Reinforcement learning on structure-conditioned categorical diffusion for protein inverse folding},
  author={Ektefaie, Yasha and Viessmann, Olivia and Narayanan, Siddharth and Dresser, Drew and Kim, J Mark and Mkrtchyan, Armen},
  journal={arXiv preprint arXiv:2410.17173},
  year={2024}
}

@article{song2024surfpro,
  title={SurfPro: Functional Protein Design Based on Continuous Surface},
  author={Song, Zhenqiao and Huang, Tinglin and Li, Lei and Jin, Wengong},
  journal={arXiv preprint arXiv:2405.06693},
  year={2024}
}

@inproceedings{gao2023kw,
  title={KW-Design: Pushing the Limit of Protein Design via Knowledge Refinement},
  author={Gao, Zhangyang and Tan, Cheng and Chen, Xingran and Zhang, Yijie and Xia, Jun and Li, Siyuan and Li, Stan Z},
  booktitle={The Twelfth International Conference on Learning Representations},
  year={2023}
}

@article{gao2022pifold,
  title={PiFold: Toward effective and efficient protein inverse folding},
  author={Gao, Zhangyang and Tan, Cheng and Chacon, Pablo and Li, Stan Z},
  journal={arXiv preprint arXiv:2209.12643},
  year={2022}
}

@inproceedings{zheng2023structure,
  title={Structure-informed language models are protein designers},
  author={Zheng, Zaixiang and Deng, Yifan and Xue, Dongyu and Zhou, Yi and Ye, Fei and Gu, Quanquan},
  booktitle={International conference on machine learning},
  pages={42317--42338},
  year={2023},
  organization={PMLR}
}

@article{dauparas2022robust,
  title={Robust deep learning--based protein sequence design using ProteinMPNN},
  author={Dauparas, Justas and Anishchenko, Ivan and Bennett, Nathaniel and Bai, Hua and Ragotte, Robert J and Milles, Lukas F and Wicky, Basile IM and Courbet, Alexis and de Haas, Rob J and Bethel, Neville and others},
  journal={Science},
  volume={378},
  number={6615},
  pages={49--56},
  year={2022},
  publisher={American Association for the Advancement of Science}
}

@article{van2024fast,
  title={Fast and accurate protein structure search with Foldseek},
  author={Van Kempen, Michel and Kim, Stephanie S and Tumescheit, Charlotte and Mirdita, Milot and Lee, Jeongjae and Gilchrist, Cameron LM and Soding, Johannes and Steinegger, Martin},
  journal={Nature biotechnology},
  volume={42},
  number={2},
  pages={243--246},
  year={2024},
  publisher={Nature Publishing Group US New York}
}

@article{somnath2021multi,
  title={Multi-scale representation learning on proteins},
  author={Somnath, Vignesh Ram and Bunne, Charlotte and Krause, Andreas},
  journal={Advances in Neural Information Processing Systems},
  volume={34},
  pages={25244--25255},
  year={2021}
}

@inproceedings{lee2023pre,
  title={Pre-training Sequence, Structure, and Surface Features for Comprehensive Protein Representation Learning},
  author={Lee, Youhan and Yu, Hasun and Lee, Jaemyung and Kim, Jaehoon},
  booktitle={The Twelfth International Conference on Learning Representations},
  year={2023}
}

@article{gainza2020deciphering,
  title={Deciphering interaction fingerprints from protein molecular surfaces using geometric deep learning},
  author={Gainza, Pablo and Sverrisson, Freyr and Monti, Frederico and Rodola, Emanuele and Boscaini, Davide and Bronstein, Michael M and Correia, Bruno E},
  journal={Nature Methods},
  volume={17},
  number={2},
  pages={184--192},
  year={2020},
  publisher={Nature Publishing Group US New York}
}

@inproceedings{sverrisson2021fast,
  title={Fast end-to-end learning on protein surfaces},
  author={Sverrisson, Freyr and Feydy, Jean and Correia, Bruno E and Bronstein, Michael M},
  booktitle={Proceedings of the IEEE/CVF Conference on Computer Vision and Pattern Recognition},
  pages={15272--15281},
  year={2021}
}

@article{connolly1983analytical,
  title={Analytical molecular surface calculation},
  author={Connolly, Michael L},
  journal={Journal of applied crystallography},
  volume={16},
  number={5},
  pages={548--558},
  year={1983},
  publisher={International Union of Crystallography}
}

@article{su2023saprot,
  title={Saprot: Protein language modeling with structure-aware vocabulary},
  author={Su, Jin and Han, Chenchen and Zhou, Yuyang and Shan, Junjie and Zhou, Xibin and Yuan, Fajie},
  journal={bioRxiv},
  pages={2023--10},
  year={2023},
  publisher={Cold Spring Harbor Laboratory}
}

@article{kabsch1983dictionary,
  title={Dictionary of protein secondary structure: pattern recognition of hydrogen-bonded and geometrical features},
  author={Kabsch, Wolfgang and Sander, Christian},
  journal={Biopolymers: Original Research on Biomolecules},
  volume={22},
  number={12},
  pages={2577--2637},
  year={1983},
  publisher={Wiley Online Library}
}

@article{tan2024simple,
  title={Simple, efficient and scalable structure-aware adapter boosts protein language models},
  author={Tan, Yang and Li, Mingchen and Zhou, Bingxin and Zhong, Bozitao and Zheng, Lirong and Tan, Pan and Zhou, Ziyi and Yu, Huiqun and Fan, Guisheng and Hong, Liang},
  journal={arXiv preprint arXiv:2404.14850},
  year={2024}
}

@article{sledzieski2024democratizing,
  title={Democratizing protein language models with parameter-efficient fine-tuning},
  author={Sledzieski, Samuel and Kshirsagar, Meghana and Baek, Minkyung and Dodhia, Rahul and Lavista Ferres, Juan and Berger, Bonnie},
  journal={Proceedings of the National Academy of Sciences},
  volume={121},
  number={26},
  pages={e2405840121},
  year={2024},
  publisher={National Acad Sciences}
}

@article{hu2021lora,
  title={Lora: Low-rank adaptation of large language models},
  author={Hu, Edward J and Shen, Yelong and Wallis, Phillip and Allen-Zhu, Zeyuan and Li, Yuanzhi and Wang, Shean and Wang, Lu and Chen, Weizhu},
  journal={arXiv preprint arXiv:2106.09685},
  year={2021}
}

@article{lester2021power,
  title={The power of scale for parameter-efficient prompt tuning},
  author={Lester, Brian and Al-Rfou, Rami and Constant, Noah},
  journal={arXiv preprint arXiv:2104.08691},
  year={2021}
}

@article{nijkamp2023progen2,
  title={Progen2: exploring the boundaries of protein language models},
  author={Nijkamp, Erik and Ruffolo, Jeffrey A and Weinstein, Eli N and Naik, Nikhil and Madani, Ali},
  journal={Cell systems},
  volume={14},
  number={11},
  pages={968--978},
  year={2023},
  publisher={Elsevier}
}

@article{chen2019improve,
  title={To improve protein sequence profile prediction through image captioning on pairwise residue distance map},
  author={Chen, Sheng and Sun, Zhe and Lin, Lihua and Liu, Zifeng and Liu, Xun and Chong, Yutian and Lu, Yutong and Zhao, Huiying and Yang, Yuedong},
  journal={Journal of chemical information and modeling},
  volume={60},
  number={1},
  pages={391--399},
  year={2019},
  publisher={ACS Publications}
}

@article{wang2018computational,
  title={Computational protein design with deep learning neural networks},
  author={Wang, Jingxue and Cao, Huali and Zhang, John ZH and Qi, Yifei},
  journal={Scientific reports},
  volume={8},
  number={1},
  pages={1--9},
  year={2018},
  publisher={Nature Publishing Group}
}

@article{o2018spin2,
  title={SPIN2: Predicting sequence profiles from protein structures using deep neural networks},
  author={O'Connell, James and Li, Zhixiu and Hanson, Jack and Heffernan, Rhys and Lyons, James and Paliwal, Kuldip and Dehzangi, Abdollah and Yang, Yuedong and Zhou, Yaoqi},
  journal={Proteins: Structure, Function, and Bioinformatics},
  volume={86},
  number={6},
  pages={629--633},
  year={2018},
  publisher={Wiley Online Library}
}

@article{li20143d,
  title={3D representations of amino acids—applications to protein sequence comparison and classification},
  author={Li, Jie and Koehl, Patrice},
  journal={Computational and structural biotechnology journal},
  volume={11},
  number={18},
  pages={47--58},
  year={2014},
  publisher={Elsevier}
}

@article{zhang2022ontoprotein,
  title={Ontoprotein: Protein pretraining with gene ontology embedding},
  author={Zhang, Ningyu and Bi, Zhen and Liang, Xiaozhuan and Cheng, Siyuan and Hong, Haosen and Deng, Shumin and Lian, Jiazhang and Zhang, Qiang and Chen, Huajun},
  journal={arXiv preprint arXiv:2201.11147},
  year={2022}
}

@article{orengo1997cath,
  title={CATH--a hierarchic classification of protein domain structures},
  author={Orengo, Christine A and Michie, Alex D and Jones, Susan and Jones, David T and Swindells, Mark B and Thornton, Janet M},
  journal={Structure},
  volume={5},
  number={8},
  pages={1093--1109},
  year={1997},
  publisher={Elsevier}
}

@article{li2014direct,
  title={Direct prediction of profiles of sequences compatible with a protein structure by neural networks with fragment-based local and energy-based nonlocal profiles},
  author={Li, Zhixiu and Yang, Yuedong and Faraggi, Eshel and Zhan, Jian and Zhou, Yaoqi},
  journal={Proteins: Structure, Function, and Bioinformatics},
  volume={82},
  number={10},
  pages={2565--2573},
  year={2014},
  publisher={Wiley Online Library}
}

@article{qi2020densecpd,
  title={DenseCPD: improving the accuracy of neural-network-based computational protein sequence design with DenseNet},
  author={Qi, Yifei and Zhang, John ZH},
  journal={Journal of chemical information and modeling},
  volume={60},
  number={3},
  pages={1245--1252},
  year={2020},
  publisher={ACS Publications}
}

@article{mao2023modeling,
  title={Modeling protein structure using geometric vector field networks},
  author={Mao, Weian and Zhu, Muzhi and Chen, Hao and Shen, Chunhua},
  journal={bioRxiv},
  pages={2023--05},
  year={2023},
  publisher={Cold Spring Harbor Laboratory}
}

@inproceedings{tan2023global,
  title={Global-context aware generative protein design},
  author={Tan, Cheng and Gao, Zhangyang and Xia, Jun and Hu, Bozhen and Li, Stan Z},
  booktitle={ICASSP 2023-2023 IEEE International Conference on Acoustics, Speech and Signal Processing (ICASSP)},
  pages={1--5},
  year={2023},
  organization={IEEE}
}

@article{gao2022alphadesign,
  title={Alphadesign: A graph protein design method and benchmark on alphafolddb},
  author={Gao, Zhangyang and Tan, Cheng and Li, Stan Z},
  journal={arXiv preprint arXiv:2202.01079},
  year={2022}
}

@article{wang2024diffusion,
  title={Diffusion Language Models Are Versatile Protein Learners},
  author={Wang, Xinyou and Zheng, Zaixiang and Ye, Fei and Xue, Dongyu and Huang, Shujian and Gu, Quanquan},
  journal={arXiv preprint arXiv:2402.18567},
  year={2024}
}

@inproceedings{wudiffantiseq,
  title={DiffAntiSeq: A Controllable Diffusion Model for Efficient Antibody Library Design},
  author={Wu, Fang},
  booktitle={LLM for Scientific Discovery: Reasoning, Assistance, and Collaboration},
  year={2025}
}

@inproceedings{tang2025bc,
  title={BC-DESIGN: A Biochemistry-Aware Framework for Highly Accurate Inverse Protein Folding},
  author={Tang, Xiangru and Ye, Xinwu and Wu, Fang and Shao, Daniel and Xu, Dong and Gerstein, Mark},
  booktitle={ICML 2025 Generative AI and Biology (GenBio) Workshop},
  year={2025}
}

@article{wu2026semi,
  title={A semi-supervised molecular learning framework for activity cliff estimation},
  author={Wu, Fang},
  journal={arXiv preprint arXiv:2601.04507},
  year={2026}
}

@inproceedings{wu2024semi,
  title={A semi-supervised molecular learning framework for activity cliff estimation},
  author={Wu, Fang},
  booktitle={Proceedings of the Thirty-Third International Joint Conference on Artificial Intelligence},
  pages={6080--6088},
  year={2024}
}

@article{qiu2024instructplm,
  title={InstructPLM: Aligning Protein Language Models to Follow Protein Structure Instructions},
  author={Qiu, Jiezhong and Xu, Junde and Hu, Jie and Cao, Hanqun and Hou, Liya and Gao, Zijun and Zhou, Xinyi and Li, Anni and Li, Xiujuan and Cui, Bin and others},
  journal={bioRxiv},
  pages={2024--04},
  year={2024},
  publisher={Cold Spring Harbor Laboratory}
}

@article{jumper2021highly,
  title={Highly accurate protein structure prediction with AlphaFold},
  author={Jumper, John and Evans, Richard and Pritzel, Alexander and Green, Tim and Figurnov, Michael and Ronneberger, Olaf and Tunyasuvunakool, Kathryn and Bates, Russ and Zidek, Augustin and Potapenko, Anna and others},
  journal={Nature},
  volume={596},
  number={7873},
  pages={583--589},
  year={2021},
  publisher={Nature Publishing Group}
}

@article{varadi2022alphafold,
  title={AlphaFold Protein Structure Database: massively expanding the structural coverage of protein-sequence space with high-accuracy models},
  author={Varadi, Mihaly and Anyango, Stephen and Deshpande, Mandar and Nair, Sreenath and Natassia, Cindy and Yordanova, Galabina and Yuan, David and Stroe, Oana and Wood, Gemma and Laydon, Agata and others},
  journal={Nucleic acids research},
  volume={50},
  number={D1},
  pages={D439--D444},
  year={2022},
  publisher={Oxford University Press}
}

@article{wu2022pre,
  title={Pre-Training of Equivariant Graph Matching Networks with Conformation Flexibility for Drug Binding},
  author={Wu, Fang and Jin, Shuting and Jiang, Yinghui and Jin, Xurui and Tang, Bowen and Niu, Zhangming and Liu, Xiangrong and Zhang, Qiang and Zeng, Xiangxiang and Li, Stan Z},
  journal={Advanced Science},
  volume={9},
  number={33},
  pages={2203796},
  year={2022},
  publisher={Wiley Online Library}
}

@article{rives2021biological,
  title={Biological structure and function emerge from scaling unsupervised learning to 250 million protein sequences},
  author={Rives, Alexander and Meier, Joshua and Sercu, Tom and Goyal, Siddharth and Lin, Zeming and Liu, Jason and Guo, Demi and Ott, Myle and Zitnick, C Lawrence and Ma, Jerry and others},
  journal={Proceedings of the National Academy of Sciences},
  volume={118},
  number={15},
  pages={e2016239118},
  year={2021},
  publisher={National Acad Sciences}
}

@article{rao2019evaluating,
  title={Evaluating protein transfer learning with TAPE},
  author={Rao, Roshan and Bhattacharya, Nicholas and Thomas, Neil and Duan, Yan and Chen, Peter and Canny, John and Abbeel, Pieter and Song, Yun},
  journal={Advances in neural information processing systems},
  volume={32},
  year={2019}
}

@article{elnaggar2020prottrans,
  title={ProtTrans: towards cracking the language of Life's code through self-supervised deep learning and high performance computing},
  author={Elnaggar, Ahmed and Heinzinger, Michael and Dallago, Christian and Rihawi, Ghalia and Wang, Yu and Jones, Llion and Gibbs, Tom and Feher, Tamas and Angerer, Christoph and Steinegger, Martin and others},
  journal={arXiv preprint arXiv:2007.06225},
  year={2020}
}

@article{lin2022language,
  title={Language models of protein sequences at the scale of evolution enable accurate structure prediction},
  author={Lin, Zeming and Akin, Halil and Rao, Roshan and Hie, Brian and Zhu, Zhongkai and Lu, Wenting and dos Santos Costa, Allan and Fazel-Zarandi, Maryam and Sercu, Tom and Candido, Sal and others},
  journal={bioRxiv},
  year={2022},
  publisher={Cold Spring Harbor Laboratory}
}

@article{jing2020learning,
  title={Learning from protein structure with geometric vector perceptrons},
  author={Jing, Bowen and Eismann, Stephan and Suriana, Patricia and Townshend, Raphael JL and Dror, Ron},
  journal={arXiv preprint arXiv:2009.01411},
  year={2020}
}

@inproceedings{satorras2021n,
  title={E (n) equivariant graph neural networks},
  author={Satorras, Victor Garcia and Hoogeboom, Emiel and Welling, Max},
  booktitle={International conference on machine learning},
  pages={9323--9332},
  year={2021},
  organization={PMLR}
}

@article{ingraham2019generative,
  title={Generative models for graph-based protein design},
  author={Ingraham, John and Garg, Vikas and Barzilay, Regina and Jaakkola, Tommi},
  journal={Advances in neural information processing systems},
  volume={32},
  year={2019}
}

@article{zhang2022protein,
  title={Protein representation learning by geometric structure pretraining},
  author={Zhang, Zuobai and Xu, Minghao and Jamasb, Arian and Chenthamarakshan, Vijil and Lozano, Aurelie and Das, Payel and Tang, Jian},
  journal={arXiv preprint arXiv:2203.06125},
  year={2022}
}

@article{hsu2022learning,
  title={Learning inverse folding from millions of predicted structures},
  author={Hsu, Chloe and Verkuil, Robert and Liu, Jason and Lin, Zeming and Hie, Brian and Sercu, Tom and Lerer, Adam and Rives, Alexander},
  journal={bioRxiv},
  year={2022},
  publisher={Cold Spring Harbor Laboratory}
}

@article{strokach2020fast,
  title={Fast and flexible protein design using deep graph neural networks},
  author={Strokach, Alexey and Becerra, David and Corbi-Verge, Carles and Perez-Riba, Albert and Kim, Philip M},
  journal={Cell systems},
  volume={11},
  number={4},
  pages={402--411},
  year={2020},
  publisher={Elsevier}
}

@article{zhang2004scoring,
  title={Scoring function for automated assessment of protein structure template quality},
  author={Zhang, Yang and Skolnick, Jeffrey},
  journal={Proteins: Structure, Function, and Bioinformatics},
  volume={57},
  number={4},
  pages={702--710},
  year={2004},
  publisher={Wiley Online Library}
}

@article{zhang2008curvature,
  title={Curvature estimation of 3D point cloud surfaces through the fitting of normal section curvatures},
  author={Zhang, Xiaopeng and Li, Hongjun and Cheng, Zhanglin and others},
  journal={Proceedings of ASIAGRAPH},
  volume={2008},
  number={23-26},
  pages={2},
  year={2008}
}

@article{yim2023se,
  title={SE (3) diffusion model with application to protein backbone generation},
  author={Yim, Jason and Trippe, Brian L and De Bortoli, Valentin and Mathieu, Emile and Doucet, Arnaud and Barzilay, Regina and Jaakkola, Tommi},
  journal={arXiv preprint arXiv:2302.02277},
  year={2023}
}

@article{abramson2024accurate,
  title={Accurate structure prediction of biomolecular interactions with AlphaFold 3},
  author={Abramson, Josh and Adler, Jonas and Dunger, Jack and Evans, Richard and Green, Tim and Pritzel, Alexander and Ronneberger, Olaf and Willmore, Lindsay and Ballard, Andrew J and Bambrick, Joshua and others},
  journal={Nature},
  pages={1--3},
  year={2024},
  publisher={Nature Publishing Group UK London}
}

@article{savinov2021step,
  title={Step-unrolled denoising autoencoders for text generation},
  author={Savinov, Nikolay and Chung, Junyoung and Binkowski, Mikolaj and Elsen, Erich and Oord, Aaron van den},
  journal={arXiv preprint arXiv:2112.06749},
  year={2021}
}

@inproceedings{alexa2001point,
  title={Point set surfaces},
  author={Alexa, Marc and Behr, Johannes and Cohen-Or, Daniel and Fleishman, Shachar and Levin, David and Silva, Claudio T},
  booktitle={Proceedings Visualization, 2001. VIS'01.},
  pages={21--29},
  year={2001},
  organization={IEEE}
}

@article{berman2002protein,
  title={The protein data bank},
  author={Berman, Helen M and Battistuz, Tammy and Bhat, Talapady N and Bluhm, Wolfgang F and Bourne, Philip E and Burkhardt, Kyle and Feng, Zukang and Gilliland, Gary L and Iype, Lisa and Jain, Shri and others},
  journal={Acta Crystallographica Section D: Biological Crystallography},
  volume={58},
  number={6},
  pages={899--907},
  year={2002},
  publisher={International Union of Crystallography}
}

@article{hu2022exploring,
  title={Exploring evolution-aware \&-free protein language models as protein function predictors},
  author={Hu, Mingyang and Yuan, Fajie and Yang, Kevin and Ju, Fusong and Su, Jin and Wang, Hui and Yang, Fei and Ding, Qiuyang},
  journal={Advances in Neural Information Processing Systems},
  volume={35},
  pages={38873--38884},
  year={2022}
}

@article{bennett2023improving,
  title={Improving de novo protein binder design with deep learning},
  author={Bennett, Nathaniel R and Coventry, Brian and Goreshnik, Inna and Huang, Buwei and Allen, Aza and Vafeados, Dionne and Peng, Ying Po and Dauparas, Justas and Baek, Minkyung and Stewart, Lance and others},
  journal={Nature Communications},
  volume={14},
  number={1},
  pages={2625},
  year={2023},
  publisher={Nature Publishing Group UK London}
}

@article{defresne2021protein,
  title={Protein design with deep learning},
  author={Defresne, Marianne and Barbe, Sophie and Schiex, Thomas},
  journal={International Journal of Molecular Sciences},
  volume={22},
  number={21},
  pages={11741},
  year={2021},
  publisher={MDPI}
}

@article{wu2024hierarchical,
  title={A hierarchical training paradigm for antibody structure-sequence co-design},
  author={Wu, Fang and Li, Stan Z},
  journal={Advances in Neural Information Processing Systems},
  volume={36},
  year={2024}
}

@article{nwankpa2018activation,
  title={Activation functions: Comparison of trends in practice and research for deep learning},
  author={Nwankpa, Chigozie and Ijomah, Winifred and Gachagan, Anthony and Marshall, Stephen},
  journal={arXiv preprint arXiv:1811.03378},
  year={2018}
}

@inproceedings{mitra2003estimating,
  title={Estimating surface normals in noisy point cloud data},
  author={Mitra, Niloy J and Nguyen, An},
  booktitle={Proceedings of the nineteenth annual symposium on Computational geometry},
  pages={322--328},
  year={2003}
}

@inproceedings{tian2023geomae,
  title={Geomae: Masked geometric target prediction for self-supervised point cloud pre-training},
  author={Tian, Xiaoyu and Ran, Haoxi and Wang, Yue and Zhao, Hang},
  booktitle={Proceedings of the IEEE/CVF Conference on Computer Vision and Pattern Recognition},
  pages={13570--13580},
  year={2023}
}

@book{kobayashi1996foundations,
  title={Foundations of differential geometry, volume 2},
  author={Kobayashi, Shoshichi and Nomizu, Katsumi},
  volume={61},
  year={1996},
  publisher={John Wiley \& Sons}
}

@article{gao2024uniif,
  title={Uniif: Unified molecule inverse folding},
  author={Gao, Zhangyang and Wang, Jue and Tan, Cheng and Wu, Lirong and Huang, Yufei and Li, Siyuan and Ye, Zhirui and Li, Stan Z},
  journal={arXiv preprint arXiv:2405.18968},
  year={2024}
}

@article{gasteiger2020directional,
  title={Directional message passing for molecular graphs},
  author={Gasteiger, Johannes and Gross, Janek and Gunnemann, Stephan},
  journal={arXiv preprint arXiv:2003.03123},
  year={2020}
}

@article{gasteiger2020fast,
  title={Fast and uncertainty-aware directional message passing for non-equilibrium molecules},
  author={Gasteiger, Johannes and Giri, Shankari and Margraf, Johannes T and Gunnemann, Stephan},
  journal={arXiv preprint arXiv:2011.14115},
  year={2020}
}

@article{devlin2018bert,
  title={Bert: Pre-training of deep bidirectional transformers for language understanding},
  author={Devlin, Jacob and Chang, Ming-Wei and Lee, Kenton and Toutanova, Kristina},
  journal={arXiv preprint arXiv:1810.04805},
  year={2018}
}

@article{zhang2023equipocket,
  title={Equipocket: an e (3)-equivariant geometric graph neural network for ligand binding site prediction},
  author={Zhang, Yang and Huang, Wenbing and Wei, Zhewei and Yuan, Ye and Ding, Zhaohan},
  journal={arXiv preprint arXiv:2302.12177},
  year={2023}
}

@article{pearson2005limits,
  title={The limits of protein sequence comparison?},
  author={Pearson, William R and Sierk, Michael L},
  journal={Current opinion in structural biology},
  volume={15},
  number={3},
  pages={254--260},
  year={2005},
  publisher={Elsevier}
}

@article{cock2009biopython,
  title={Biopython: freely available Python tools for computational molecular biology and bioinformatics},
  author={Cock, Peter JA and Antao, Tiago and Chang, Jeffrey T and Chapman, Brad A and Cox, Cymon J and Dalke, Andrew and Friedberg, Iddo and Hamelryck, Thomas and Kauff, Frank and Wilczynski, Bartek and others},
  journal={Bioinformatics},
  volume={25},
  number={11},
  pages={1422},
  year={2009},
  publisher={Oxford University Press}
}

@article{lai2024interformer,
  title={Interformer: An Interaction-Aware Model for Protein-Ligand Docking and Affinity Prediction},
  author={Lai, Houtim and Wang, Longyue and Qian, Ruiyuan and Ye, Geyan and Huang, Juhong and Wu, Fandi and Wu, Fang and Zeng, Xiangxiang and Liu, Wei},
  year={2024}
}

@article{wang2023pdb,
  title={PDB-Struct: A Comprehensive Benchmark for Structure-based Protein Design},
  author={Wang, Chuanrui and Zhong, Bozitao and Zhang, Zuobai and Chaudhary, Narendra and Misra, Sanchit and Tang, Jian},
  journal={arXiv preprint arXiv:2312.00080},
  year={2023}
}

@article{tunyasuvunakool2021highly,
  title={Highly accurate protein structure prediction for the human proteome},
  author={Tunyasuvunakool, Kathryn and Adler, Jonas and Wu, Zachary and Green, Tim and Zielinski, Michal and {\v{Z}}{\'\i}dek, Augustin and Bridgland, Alex and Cowie, Andrew and Meyer, Clemens and Laydon, Agata and others},
  journal={Nature},
  volume={596},
  number={7873},
  pages={590--596},
  year={2021},
  publisher={Nature Publishing Group UK London}
}

@article{wu2022high,
  title={High-resolution de novo structure prediction from primary sequence},
  author={Wu, Ruidong and Ding, Fan and Wang, Rui and Shen, Rui and Zhang, Xiwen and Luo, Shitong and Su, Chenpeng and Wu, Zuofan and Xie, Qi and Berger, Bonnie and others},
  journal={BioRxiv},
  pages={2022--07},
  year={2022},
  publisher={Cold Spring Harbor Laboratory}
}

@article{robinson2014msm,
  title={MSM: a new flexible framework for multimodal surface matching},
  author={Robinson, Emma C and Jbabdi, Saad and Glasser, Matthew F and Andersson, Jesper and Burgess, Gregory C and Harms, Michael P and Smith, Stephen M and Van Essen, David C and Jenkinson, Mark},
  journal={Neuroimage},
  volume={100},
  pages={414--426},
  year={2014},
  publisher={Elsevier}
}

@article{delano2002pymol,
  title={Pymol: An open-source molecular graphics tool},
  author={DeLano, Warren L and others},
  journal={CCP4 Newsl. Protein Crystallogr},
  volume={40},
  number={1},
  pages={82--92},
  year={2002},
  publisher={Citeseer}
}

@inproceedings{wu2024surface,
  title={Surface-VQMAE: Vector-quantized Masked Auto-encoders on Molecular Surfaces},
  author={Wu, Fang and Li, Stan Z},
  booktitle={International Conference on Machine Learning},
  pages={53619--53634},
  year={2024},
  organization={PMLR}
}

@article{sun2024dsr,
  title={DSR: dynamical surface representation as implicit neural networks for protein},
  author={Sun, Daiwen and Huang, He and Li, Yao and Gong, Xinqi and Ye, Qiwei},
  journal={Advances in Neural Information Processing Systems},
  volume={36},
  year={2024}
}

@inproceedings{park2019deepsdf,
  title={Deepsdf: Learning continuous signed distance functions for shape representation},
  author={Park, Jeong Joon and Florence, Peter and Straub, Julian and Newcombe, Richard and Lovegrove, Steven},
  booktitle={Proceedings of the IEEE/CVF conference on computer vision and pattern recognition},
  pages={165--174},
  year={2019}
}

@article{gainza2023novo,
  title={De novo design of protein interactions with learned surface fingerprints},
  author={Gainza, Pablo and Wehrle, Sarah and Van Hall-Beauvais, Alexandra and Marchand, Anthony and Scheck, Andreas and Harteveld, Zander and Buckley, Stephen and Ni, Dongchun and Tan, Shuguang and Sverrisson, Freyr and others},
  journal={Nature},
  volume={617},
  number={7959},
  pages={176--184},
  year={2023},
  publisher={Nature Publishing Group UK London}
}

\newpage
\appendix
\onecolumn

\section{Mathematical Analysis}
\label{app:curvature}

We show that the curvature feature $\boldsymbol{\psi}$ is roto-translation invariant.

For a translation $\mathbf{x}_j'=\mathbf{x}_j+\mathbf{t}$ with $\mathbf{t}\in\mathbb{R}^3$, the centroid becomes $\bar{\mathbf{x}}'=\bar{\mathbf{x}}+\mathbf{t}$. The translated covariance matrix is 
$\boldsymbol{\Sigma}'=\frac{1}{|\mathcal{N}_{(i)}|}\sum_{\mathbf{x}_j\in\mathcal{N}_{(i)}}(\mathbf{x}_j+\mathbf{t})(\mathbf{x}_j+\mathbf{t})^\top-\bar{\mathbf{x}}'\bar{\mathbf{x}}'^\top$. Expanding and canceling translation terms yields 
$\boldsymbol{\Sigma}'=\frac{1}{|\mathcal{N}_{(i)}|}\sum_{\mathbf{x}_j\in\mathcal{N}_{(i)}}\mathbf{x}_j\mathbf{x}_j^\top-\bar{\mathbf{x}}\bar{\mathbf{x}}^\top=\boldsymbol{\Sigma}$, proving translation invariance.

For a rotation $\mathbf{R}\in\mathrm{SO}(3)$, let $\mathbf{x}_j'=\mathbf{R}\mathbf{x}_j$ and $\bar{\mathbf{x}}'=\mathbf{R}\bar{\mathbf{x}}$. Then,
$\boldsymbol{\Sigma}'=\frac{1}{|\mathcal{N}_{(i)}|}\sum_{\mathbf{x}_j\in\mathcal{N}_{(i)}}\mathbf{R}\mathbf{x}_j(\mathbf{R}\mathbf{x}_j)^\top-\bar{\mathbf{x}}'\bar{\mathbf{x}}'^\top=\mathbf{R}\boldsymbol{\Sigma}\mathbf{R}^\top$. Since $\mathbf{R}\boldsymbol{\Sigma}\mathbf{R}^\top$ is similar to $\boldsymbol{\Sigma}$, they share the same eigenvalues $\epsilon_1,\epsilon_2,\epsilon_3$. Therefore, $\boldsymbol{\psi}$ is invariant under rotations and translations.

\section{Experimental Details}
\label{app:exp_details}
\paragraph{Training and metrics.} The models were trained for 50 epochs by default, using the Adam optimizer on 4 A100 GPUs. We used the same training settings as ProteinMPNN~\citep{dauparas2022robust} and LM-Design, with a batch size of approximately 6000 residues and the Adam optimizer configured with a \textsc{noam} learning rate scheduler. Following previous works, perplexity and \emph{median} AAR scores are reported. In Tab.~\ref{tab:cath2} and~\ref{tab:cath3}, two subsets of the entire test set are also reported. Particularly, the \textsc{Short} set contains proteins up to length 100, and the \textsc{Single chain} set contains proteins recorded as a single chain in PDB~\citep{berman2002protein}.

\textbf{Protein Binder Design.}
We evaluate SurfDesign on a curated benchmark of experimentally validated protein-protein binding complexes spanning multiple target categories. For target categories with sufficient data, complexes are split into training, validation, and test sets using an $8{:}1{:}1$ ratio; categories with limited data are evaluated in a zero-shot setting and used only for testing. All models are fine-tuned on the same training split to ensure fair comparison.

Given a target protein structure, SurfDesign generates binder sequences conditioned on the target surface geometry. We report two complementary evaluation settings. 
(i) \emph{Greedy decoding}, where a single binder sequence is generated per target and evaluated directly. 
(ii) \emph{Stochastic sampling}, where we sample $K=10$ binder sequences per target using a softmax temperature of $0.1$.

\textbf{Functional binding quality} is assessed using the AlphaFold2-predicted aligned error between the binder and target chains (pAE$_\text{interaction}$), which has been shown to correlate with binding likelihood. Lower pAE$_\text{interaction}$ indicates stronger predicted binding. For each designed binder, we first predict its monomer structure using ESMFold, superimpose it onto the binder chain position of the native complex, and then compute pAE$_\text{interaction}$ using AlphaFold2. A designed binder is considered \emph{successful} if its pAE$_\text{interaction}$ is lower than that of the corresponding native positive binder. We additionally report randomly sampled sequences of matched length as a negative control.

\paragraph{Enzyme Design.} We evaluate enzyme design using a benchmark comprising multiple enzyme classes, each paired with a specific substrate. To avoid potential data leakage from inverse folding pretraining corpora, all enzymes that overlap with the pretraining datasets are removed prior to data splitting and evaluation.

For enzyme classes with more than 100 samples, we perform clustering-based splitting followed by an $8{:}1{:}1$ train/validation/test partition. Smaller enzyme classes are evaluated exclusively in a zero-shot setting. All methods are fine-tuned on the same training split when available.

Enzyme functionality is evaluated using the Enzyme--Substrate Potential (ESP) score, which quantifies the compatibility between a designed enzyme structure and its substrate; lower ESP indicates more favorable interactions. As in binder design, we report results for both greedy decoding and stochastic sampling ($K=10$, temperature=0.1). A generated enzyme sequence is considered \emph{successful} if its ESP score improves upon that of the native enzyme for the same substrate.

\paragraph{Implementation for Surface Generation.} PyMol is used to generate surfaces in our implementation. We have tried the fast sampling algorithm introduced by dMaSIF~\citep{sverrisson2021fast} and used by later studies~\citep{wu2024surface}, which approximates the protein surface as the level set of a smooth distance function. However, this sampling mechanism exhibits unacceptable randomness and is therefore abandoned in favor of SurfDesign. 
As for the biochemical feature computation, we follow MaSIF~\citep{gainza2020deciphering} and calculate three key invariant point inputs, including the Poisson Boltzmann electrostatics using APBS~\footnote{\url{https://github.com/LPDI-EPFL/masif/blob/master/source/triangulation/computeAPBS.py}}, the hydrophobicity~\footnote{\url{https://github.com/LPDI-EPFL/masif/blob/master/source/triangulation/computeHydrophobicity.py}}, and the free electrons/protons~\footnote{\url{https://github.com/LPDI-EPFL/masif/blob/master/source/triangulation/compuSurfDesignteCharges.py}}. After a further ablation study, we discover that the hydrophobicity and the charge are pivot to the performance improvement while the electrostatics is not necessary. 

For Fig.~\ref{fig:box}, we employ RSA to determine the surface and core. To be specific, residues with an RSA greater than 0.25 are considered on the surface, while residues with an RSA less than 0.1 are regarded as core residues. We use the DSSP algorithm to decide the loop regions. 


\subsection{Dataset Information}
Tab.~\ref{tab:cath_4.2_info} documents the vertex count statistics for the CATH datasets. We observe an equal distribution of vertices across different splits. Besides, comparing our surface with SurfPro~\citep{song2024surfpro}, it can be found that our surface is more sparse, with nearly half of the average number of vertices per residue. This difference is due to the different computational techniques employed by various software for surface generation (\emph{e.g.}, PyMOL and MSMS). 
\begin{table}[h]
    \caption{Vertex counts statistics for surfaces from the CATH 4.2 and CATH 4.3 datasets.}
    \centering
    \resizebox{0.65\columnwidth}{!}{%
    \begin{tabular}{c|ccc|ccc} \toprule
     \multirow{2}{*}{ \textbf{Vertex Count}}   &  \multicolumn{3}{c|}{\textbf{CATH 4.2}} & \multicolumn{3}{c}{\textbf{CATH 4.3}} \\
     & Train  &   Validation  &   Test & Train  &   Validation  &   Test   \\ \midrule
     Average Vertex Count Per Residue  & 53.47  & 53.56 & 53.31  & 53.36 & 55.27 & 53.11  \\
     Maximum Vertex Count &  27,817  & 25,614 &  25,433 &  27,110 & 27,817 & 25,968 \\
     Minimum Vertex Count &  1,923  &  2,315  & 2,022 & 1,923 &  2,011 &  2,000\\ 
     Preprocess Time Per Protein &  \multicolumn{3}{c|}{0.38s} & \multicolumn{3}{c}{0.36s} \\ \bottomrule
    \end{tabular}}
    \label{tab:cath_4.2_info}
\end{table}

\paragraph{Surface Comparison}
\label{app:surface_compare}
Motivated by DSR~\citep{sun2024dsr}, we employ IoU, CD, and NC to assess the similarity between the molecular surfaces of designed proteins and target proteins. For simplicity, these three metrics are normalized to the range $0-1$. They provide a comprehensive evaluation of the model's performance from different perspectives and are defined as follows. 

\textbf{IoU.} IoU compares the reconstructed volume with the ground truth shape (higher is better). For two arbitrary shapes $A, B \subseteq \mathbb{S} \in \mathbb{R}^{n}$ is attained by $\text{IoU} =\frac{|A \cap B|}{|A \cup B|}$. 

\textbf{CD.} CD is a standard metric to evaluate the distance between two point sets $\mathcal{X}_{1}, \mathcal{X}_{2} \subset \mathbb{R}^{n}$ (lower is better) as $\mathrm{d}_{\mathrm{C}}\left(\mathcal{X}_{1}, \mathcal{X}_{2}\right)= \\ \frac{1}{2}\left(\mathrm{~d}_{\overrightarrow{\mathrm{C}}}\left(\mathcal{X}_{1}, \mathcal{X}_{2}\right)+\mathrm{d}_{\mathrm{C}}\left(\mathcal{X}_{2}, \mathcal{X}_{1}\right)\right)$, where $\mathrm{d}_{\overrightarrow{\mathrm{C}}}\left(\mathcal{X}_{1}, \mathcal{X}_{2}\right)=\frac{1}{\left|\mathcal{X}_{1}\right|} \sum_{\boldsymbol{x}_{1} \in \mathcal{X}_{1}} \min _{\boldsymbol{x}_{2} \in \mathcal{X}_{2}}\left\|\boldsymbol{x}_{1}-\boldsymbol{x}_{2}\right\|$. 

\textbf{NC.} NC evaluates estimated surface normals (higher is better). Normal consistency between two normalized unit vectors $n_{i}$ and $n_{j}$ is defined as the dot product between the two vectors. For evaluating the surface normals, given the object surface points and normal vectors: $X_{\text {pred }}=\left\{\left(\boldsymbol{x}_{i}, \overrightarrow{n_{i}}\right)\right\}$, and the ground truth surface points and normal vectors: $X_{g t}=\left\{\left(\boldsymbol{y}_{j}, \overrightarrow{m_{j}}\right)\right\}$, the surface normal consistency between $X_{\text {pred }}$ and $X_{g t}$, denoted as $\Gamma$, is defined as:
$\Gamma\left(X_{g t}, X_{\text {pred }}\right)=\frac{1}{\left|X_{g t}\right|} \sum_{j \in\left|X_{g t}\right|}\left|\overrightarrow{n_{j}} \cdot \vec{m}_{\theta\left(\boldsymbol{y}_{j}, X_{\text {pred }}\right)}\right|$,
where $\theta\left(\boldsymbol{y}_{j}, X_{\text {pred }}:=\left\{\left(\boldsymbol{x}_{i}, \overrightarrow{n_{i}}\right)\right\}\right)=\underset{i \in\left|X_{\text {pred }}\right|}{\arg \min }\left\|\boldsymbol{y}_{j}-\boldsymbol{x}_{i}\right\|_{2}^{2}$. 




\section{Additional Results and Visualization}
\subsection{More Results of Non-functional Inverse Design} 
While the primary goal of SurfDesign is functional protein design—including protein-protein binding and enzyme-substrate interactions —we also report extended inverse folding results as a \emph{diagnostic analysis} of structural compatibility.
Importantly, these experiments are \textbf{not} intended to evaluate functional optimality, nor do they assume a unique ground-truth sequence for a given structure.

\paragraph{Purpose and interpretation.}
Inverse folding benchmarks assess whether a model can generate amino-acid sequences that are globally consistent with a specified backbone or molecular surface geometry.
As discussed in the main text, a single protein structure or surface can accommodate many valid sequences, and the experimentally observed sequence is neither unique nor necessarily optimal. Accordingly, metrics such as AAR and perplexity should be interpreted as \emph{proxies for geometric and structural consistency}, rather than as supervised label-recovery objectives.

\paragraph{Extended results.}
Tab.~\ref{tab:cath3},~\ref{tab:multi-chain} and~\ref{tab:ts50_500} report additional inverse folding results on CATH~4.3, PDB, and TS50/TS500 benchmarks under settings consistent with prior work. Across all benchmarks, SurfDesign maintains strong performance relative to both backbone-only and prior surface-conditioned baselines, achieving consistently low perplexity and high median AAR. These results indicate that incorporating manifold-aware surface representations does not compromise global fold compatibility, and in fact improves geometric conditioning compared to backbone-only models.

\paragraph{No functional supervision or privileged information.}
We emphasize that no residue identities, evolutionary information (e.g., MSAs), functional annotations, or active-site labels are used as inputs during surface generation or model training. Surface features are computed solely from atomic geometry and physicochemical attributes derived from structure. Furthermore, inverse-folding datasets are disjoint from the functional design benchmarks used for binder and enzyme evaluation, thereby preventing cross-task leakage.

\paragraph{Relation to functional design.}
The strong inverse folding performance of SurfDesign should be viewed as evidence that surface-conditioned generation preserves global structural validity, rather than as the model's primary objective. In functional design settings—where no unique target sequence exists—success is instead measured by downstream functional proxies (e.g., AF2 pAE$_{\text{interaction}}$ and ESP score). Together, these results suggest that SurfDesign achieves a favorable balance: it maintains structural compatibility while enabling more precise control over surface geometry, which is critical for functional protein design.

Overall, the extended inverse folding results support the central claim of this work: explicitly modeling molecular surfaces as continuous geometric manifolds provides a strong and reliable conditioning signal that generalizes beyond functional tasks, without introducing shortcut learning or label memorization.

\begin{table*}[ht]
\caption{Sequence design on CATH 4.3. $\dagger$: \textsc{Single-chain} in~\citet{hsu2022learning} is defined differently. }
\label{tab:cath3}
\centering
\resizebox{0.8\width}{!}{%
\begin{tabular}{l|ccc|ccc}\toprule
\multirow{2}{*}{ \textbf{Models} } & \multicolumn{3}{c|}{ \textbf{Perplexity} ($\downarrow$)} & \multicolumn{3}{c}{ \textbf{AAR} ($\uparrow$) } \\
& Short & Single-chain & All & Short & Single-chain & All \\ \midrule
GVP~\citep{hsu2022learning} & 7.68 & $^\dagger$6.12 & 6.17 & 32.60 & 39.40 & 39.20 \\
ProteinMPNN~\citep{dauparas2022robust} & 6.31 & $\,$ 6.32 & 4.85 & 40.30 & 39.02 & 48.25 \\
ESM-IF~\citep{hsu2022learning} & 8.18 & $^\dagger$6.33 & 6.44 & 31.30 & 38.50 & 38.30 \\
$\quad$ + 1.2M AF2 Data & 6.05 & $^\dagger$4.00 & 4.01 & 38.10 & 51.50 & 51.60 \\
PiFold~\citep{gao2022pifold} & 5.88 & $\,\,$5.55 & 4.47 & 42.86 & 43.69 & 50.68 \\
VFN-IF~\citep{mao2023modeling} & -- & -- & -- & 45.34 & 53.70 & 52.18 \\
UniIF~\citep{gao2024uniif} & -- & -- & -- & 45.41 & ${54.46}$ & 53.05 \\ 
LM-Design-MPNN~\citep{zheng2023structure} & 5.88 & $\,\,$5.66 & 4.19 & 45.71 & 46.15 & 56.38 \\
LM-Design-PiFold~\citep{zheng2023structure} & 5.66 & $\,\,$5.52 & 4.01 & ${46.84}$ & 48.63 & 56.63 \\
KW-Design~\citep{gao2023kw} & $\underline{5.47}$ &  $\,\,\underline{5.23}$ &  $\underline{3.49}$ &  43.86 &  45.95 &  ${60.38}$ \\
MapDiff~\citep{bai2025mask} & -- & -- & --  & $\underline{55.56}$ & $\underline{54.99}$ & $\underline{60.68}$ \\ \midrule
\grow{SurfDesign} &$\mathbf{5.08}$ & $\,\,\mathbf{4.97}$ & $\mathbf{3.12}$ & $\mathbf{66.74}$ & $\mathbf{71.30}$ & $\mathbf{72.14}$ \\ \bottomrule
\end{tabular}}
\end{table*}

\begin{table}[ht]
\caption{Performance on multi-chain protein complex dataset (\emph{i.e.}, PDB). }
\centering
\resizebox{0.5\columnwidth}{!}{%
\begin{tabular}{c|cccc} \toprule
\multirow{2}{*}{\begin{tabular}{l} \textbf{Models} \\ length \end{tabular}} & \multicolumn{4}{c}{\textbf{AAR} ($\uparrow$)} \\
 & $L<100$ & $100 \leq L<500$ & $500 \leq L<1000$ & Full \\ \midrule
 StructGNN~\citep{ingraham2019generative} & 0.41 & 0.41 & 0.42 & 0.41 \\
 GraphTrans~\citep{ingraham2019generative} & 0.40 & 0.39 & 0.40 & 0.40 \\
 GCA~\citep{tan2023global} & 0.41 & 0.41 & 0.42 & 0.41 \\
 GVP~\citep{jing2020learning} & 0.44 & 0.42 & 0.45 & 0.43 \\
 AlphaDesign~\citep{gao2022alphadesign} & 0.48 & 0.49 & 0.50 & 0.49 \\
 ProteinMPNN~\citep{dauparas2022robust} & 0.52 & 0.53 & 0.55 & 0.53 \\
 PiFold~\citep{gao2022pifold} & $\underline{0.54}$ & $\underline{0.58}$ & $\underline{0.60}$ & $\underline{0.58}$ \\
 LM-Design-MPNN~\citep{zheng2023structure} &-- & -- &-- &  0.61 \\
 LM-Design-GVP~\citep{zheng2023structure} &-- & -- &-- & 0.62\\ 
 KWDesign~\citep{gao2023kw} & 0.59 & 0.66 & 0.67 & 0.66 \\ \midrule
\grow{SurfDesign} & $\mathbf{0.74}$ & $\mathbf{0.79}$ & $\mathbf{0.82}$ & $\mathbf{0.81}$ \\ \bottomrule
\end{tabular}}
\label{tab:multi-chain}
\end{table}
\begin{table*}[t]
\caption{Performance comparison on TS50 and TS500. Following prior work, we primarily report results from models trained on CATH 4.2. Numbers in the brackets are results from models trained on CATH 4.3.} 
\label{tab:ts50_500}
\begin{center}
\resizebox{0.75\width}{!}{%
\begin{tabular}{l|ccc|ccc} \toprule
\multirow{2}{*}{\textbf{Models}} & \multicolumn{3}{c|}{ \textbf{TS50} } & \multicolumn{3}{c}{ \textbf{TS500}} \\
& Perplexity ($\downarrow$) & AAR ($\uparrow$) & Worst ($\uparrow$) & Perplexity ($\downarrow$) & AAR ($\uparrow$) & Worst ($\uparrow$) \\ \midrule 
DenseCPD~\citep{qi2020densecpd} & -- & 50.71 & -- & -- & 55.53 & -- \\ 
StructGNN~\citep{ingraham2019generative} & 5.40 & 43.89 & 26.92 & 4.98 & 45.69 & 0.05 \\
GraphTrans~\citep{ingraham2019generative} & 5.60 & 42.20 & 29.22 & 5.16 & 44.66 & 0.03 \\
GVP~\citep{jing2020learning} & 4.71 & 44.14 & 33.73 & 4.20 & 49.14 & $\underline{0.09}$ \\
GCA~\citep{tan2023global} & 5.09 & 47.02 & 28.87 & 4.72 & 47.74 & 0.03 \\
AlphaDesign~\citep{gao2022alphadesign} & 5.25 & 48.36 & 32.31 & 4.93 & 49.23 & 0.03 \\
KW-Design~\citep{gao2023kw} & 3.10 & 62.79 & 39.31 & 2.86 & 69.19 & 0.02 \\ 
VFN-IF~\citep{mao2023modeling} & 3.58 & 59.54 & -- & 3.19 & 63.65 & -- \\ 
VFN-IF-ESM~\citep{mao2023modeling} &  2.52 & $\underline{73.30}$ & --& 2.54 & $\underline{72.49}$ & -- \\
InstructPLM~\citep{qiu2024instructplm} & $\underline{2.29}$ & 67.99 & -- & 2.42 & 64.22 & -- \\ 
PRISM~\citep{mahbub2025prism} & 2.43 & 67.92 & -- & 2.43 & 67.92& --\\ 
ProteinMPNN-DPO~\citep{xu2025protein} & 4.85  & 45.91 & -- & 4.26 & 48.23 & -- \\
InstructPLM-DPO~\citep{xu2025protein} & 2.52 & 62.01 & -- & 2.17 & 66.37 & -- \\ \midrule
ProteinMPNN~\citep{dauparas2022robust} & 3.93 (3.62) & 54.43 (54.22) & 37.24 (41.18) & 3.53 (3.27) & 58.08 (57.23) & 0.03 (0.04)\\ 
PiFold~\citep{gao2022pifold} & 3.86 (3.70) & 58.72 (59.68) & 37.93 (38.14) & 3.44 (3.70) & 60.42 (59.95) & 0.03 ($\underline{0.05}$) \\
LM-Design-MPNN~\citep{zheng2023structure} & 3.82 (3.60) & 56.92 (58.13) & 35.17 (39.14) & $\underline{2.13}$ ($\underline{2.15}$) & 64.30 (63.76) & 0.04 (0.04)\\
LM-Design-PiFold~\citep{zheng2023structure} & 3.50 ($\underline{3.27}$) & 57.89 ($\underline{61.38}$) & $\underline{39.74}$ ($\underline{46.75}$) & 3.19 (3.09) & 67.78 ($\underline{66.56}$) & 0.02 (0.04) \\ \midrule
\grow{SurfDesign} & $\mathbf{2.05}$ ($\mathbf{2.03}$) & $\mathbf{82.16}$ ($\mathbf{83.44}$) & $\mathbf{41.30}$ ($\mathbf{47.81}$) & $\mathbf{1.98}$ ($\mathbf{1.96}$) & $\mathbf{84.70}$ ($\mathbf{85.12}$) & $\mathbf{0.10}$ ($\mathbf{0.08}$)\\ \bottomrule
\end{tabular}}
\end{center}
\end{table*}

\subsection{Refoldability Analysis}
\label{app:refoldability}
Following~\citet{wang2023pdb}, firstly, to assess whether the generated sequences can respect the structure condition, we evaluate the agreement of the ground truth structure with the predicted structures using the TM-score~\citep{zhang2004scoring}. We refer to this metric as Ref-TM. Furthermore, to evaluate the folding stability of the generated sequences, we compute the mean per-residue confidence estimate, pLDDT, predicted by the structure prediction models, which we refer to as Ref-pLDDT.
As pLDDT is a reliable predictor of disorder~\citep{tunyasuvunakool2021highly}, AlphaFold-2, OmegaFold~\citep{wu2022high}, and ESMFold~\citep{lin2022language} are leveraged as a structure prediction model, which helps minimize deviations due to the choice of model.

We evaluate SurfDesign on the same 82 test samples as in the CATH dataset, and the results are reported in Tab.~\ref{tab:Refoldability}. We observe that SurfDesign stands out as the leading design method across the refoldability metrics, competitive with ProteinMPNN. It achieves 0.89 Ref-TM and 89.42 Ref-pLDDT with AlphaFold-2 prediction. ProteinMPNN is slightly behind with a 0.87 Ref-TM and 87.89 Ref-pLDDT, followed by LM-Design. 
\begin{table}[th]
    \centering
    \caption{Refoldability metric and AAR metric on the CATH dataset. We employ \textbf{bold} and \underline{underline} to highlight the best and suboptimal results on each metric. We use TM and pLDDT to represent Ref-TM and Ref-pLDDT.}
    \resizebox{0.6\columnwidth}{!}{%
    \begin{tabular}{c|cc|cc|cc|c} \toprule
    \multirow[]{2}{*}{\textbf{Design method}} & \multicolumn{2}{c|}{\textbf{ESMFold}} & \multicolumn{2}{c|}{\textbf{OmegaFold}} & \multicolumn{2}{c|}{\textbf{AlphaFold-2}} & \multirow[]{2}{*}{\textbf{AAR\%}} \\
    & TM & pLDDT & TM & pLDDT & TM & pLDDT & \\ \midrule
    Wildtype & 0.80 & 74.91 & 0.75 & 78.39 & 0.90 & 89.87 & 100 \\ \midrule 
    Uniform & 0.05 & 27.68 & 0.05 & 31.53 & 0.06 & 33.68 & 5.00 \\
    Natural frequencies & 0.07 & 30.53 & 0.07 & 35.59 & 0.06 & 35.02 & 5.84 \\ \midrule
    AF-Design & 0.53 & 61.37 & 0.53 & 72.04 & 0.52 & 75.29 & 15.95 \\
    ESM-Design & 0.38 & 59.65 & 0.38 & 62.66 & 0.37 & 60.02 & 17.33 \\  \midrule 
    StructTrans & 0.72 & 68.85 & 0.64 & 70.35 & 0.79 & 80.66 & 35.89 \\
    GVP & 0.73 & 69.67 & 0.67 & 74.33 & 0.83 & 84.29 & 39.46 \\
    ProteinMPNN & \underline{0.80} & \underline{76.53} & \underline{0.76} & \textbf{80.75} & \underline{0.87} & \underline{87.89} & 41.44 \\
    PiFold & 0.71 & 67.55 & 0.64 & 70.21 & 0.82 & 82.54 & 44.86 \\
    LM-Design & 0.73 & 72.12 & 0.70 & 77.58 & 0.85 & 87.26 & \underline{51.23} \\ 
    \grow{SurfDesign} &  \textbf{0.81} & \textbf{79.35} & \textbf{0.76} & \underline{80.11} & \textbf{0.89} & \textbf{89.42} & \textbf{70.19}  \\ \bottomrule
    \end{tabular}}
    \label{tab:Refoldability}
\end{table}

In addition to the AAR rate, we have incorporated Foldable Diversity and sc-TM, as recommended by~\citet {ektefaie2024reinforcement}, to further assess the diversity and self-consistency of the generated sequences. Foldable Diversity evaluates only those sequence pairs that are structurally consistent with the input protein backbone, providing a more targeted diversity metric that avoids penalizing high-quality, diverse designs. Self-consistency TM score (sc-TM), following~\citet{trippe2022diffusion}, gauges the consistency of structural predictions for generated sequences, leveraging a fixed threshold of $T M_{\min}=0.7$ as implemented by~\citet{ektefaie2024reinforcement}. We refer to \url{https://github.com/flagshippioneering/pi-rldif} for computation, and the results are shown below. The analysis shows that SurfDesign maintains high structural consistency while exhibiting competitive diversity, outperforming other methods on foldable diversity metrics and providing substantial evidence of the model's ability to generate high-quality, diverse sequences that remain faithful to the structural constraints of input proteins.
\begin{table}[th]
\centering
\caption{Foldable diversity on CATH-all.}
\begin{tabular}{lcc} \toprule 
    Model & Foldable Diversity $\uparrow$ & sc-TM $\uparrow$ \\ \midrule
    ProteinMPNN (T=0, RD) & $20 \%$ & 0.80 \\
    ProteinMPNN (T=0.1) & $23 \%$ & 0.67 \\
    ProteinMPNN (T=0.2) & $3 \%$ & 0.30 \\
    ProteinMPNN (T=0.3) & $0.1 \%$ & 0.14 \\
    PiFold (T=0.1) & $23 \%$  & 0.72 \\
    PiFold (T=0.2) & $8 \%$ & 0.38 \\
    KWDesign (T=0.1) & $18 \%$ & 0.79 \\
    KWDesign (T=0.2) & $23 \%$ & 0.58 \\ \midrule
    \grow{SurfDesign} & $\mathbf{2 3 \%}$ & $\mathbf{0 . 8 4}$ \\ \bottomrule
\end{tabular}
\end{table}

\subsection{Structure Visualization}
\label{app:visual}
\begin{figure}[ht]
    \centering
    \includegraphics[width=0.8\linewidth]{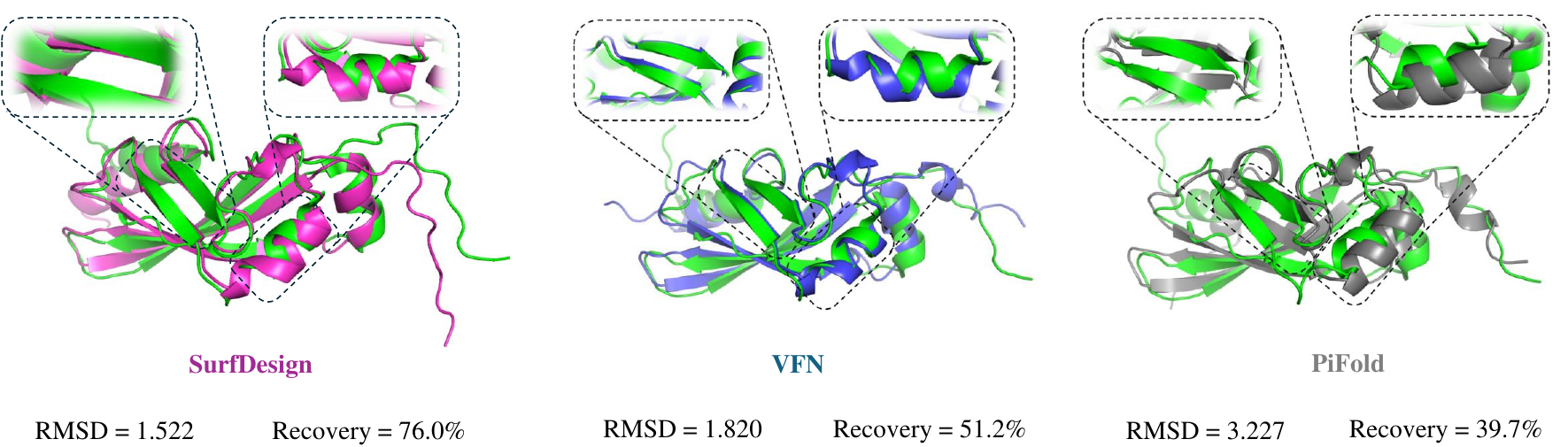}
    \caption{Visualization results of a challenging sample (PDB 2KRT). We use AlphaFold3 to recover the structure from the predicted sequence and compare it with the experimentally determined ground-truth structure.}
    \label{fig:2KRT}
\end{figure}

\begin{figure}
    \centering
    \includegraphics[width=0.8\linewidth]{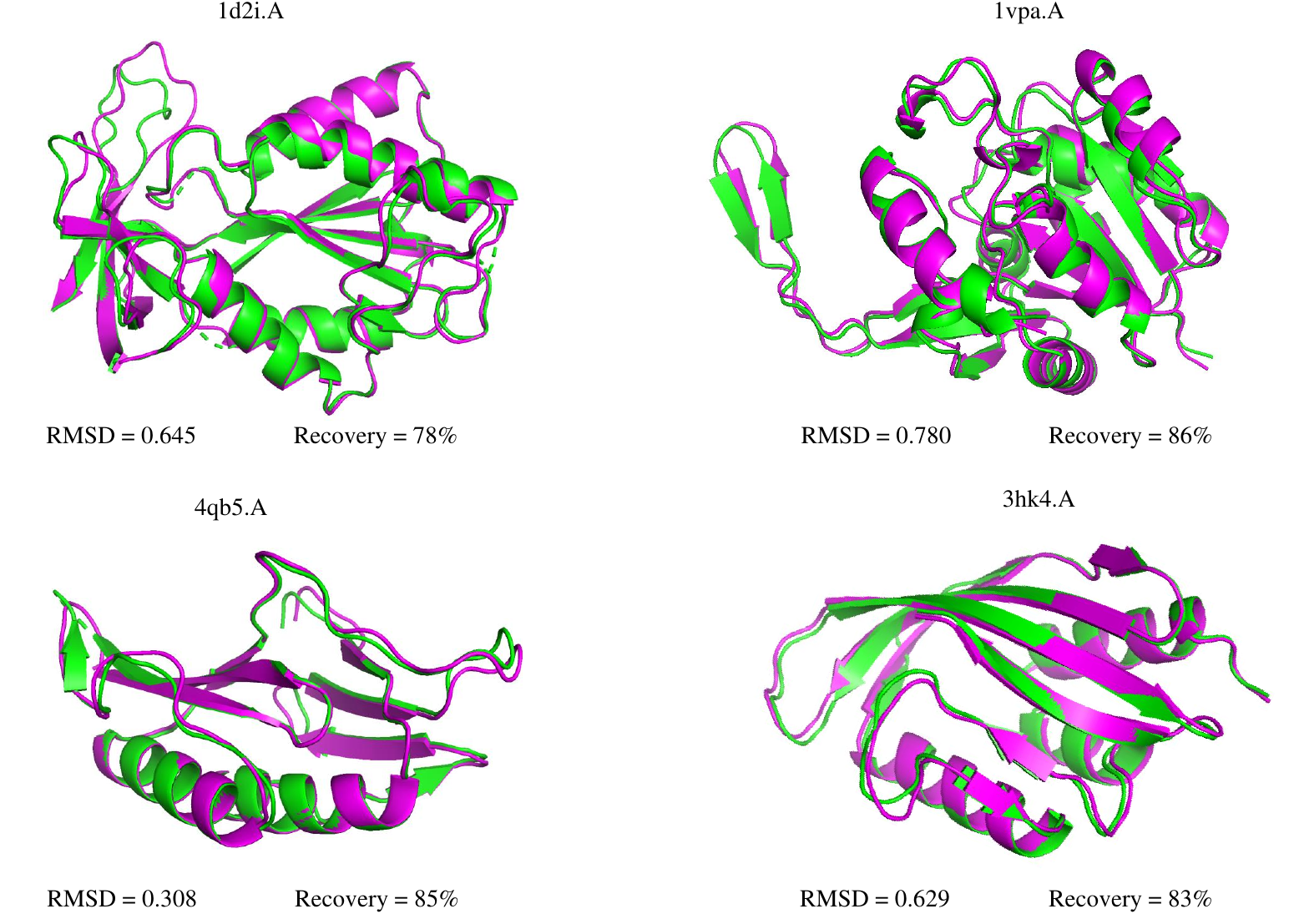}
    \caption{Visualization of SurfDesign, where the green and pink ones are ground truth and designed structures, respectively. }
    \label{fig:case}
\end{figure}

Here, we visualize several protein structure restoration results of SurfDesign, as shown in Fig.~\ref{fig:2KRT} and Fig.~\ref{fig:case}. The designed structures were obtained using the latest AlphaFold 3~\citep{abramson2024accurate}~\footnote{We employed the Alphafold Server for inference at~\url{https://alphafoldserver.com/}.}
 
\subsection{Surface Visualization}
\label{app:surf_visual}
\begin{figure}[ht]
     \centering
     \includegraphics[width=0.95\linewidth]{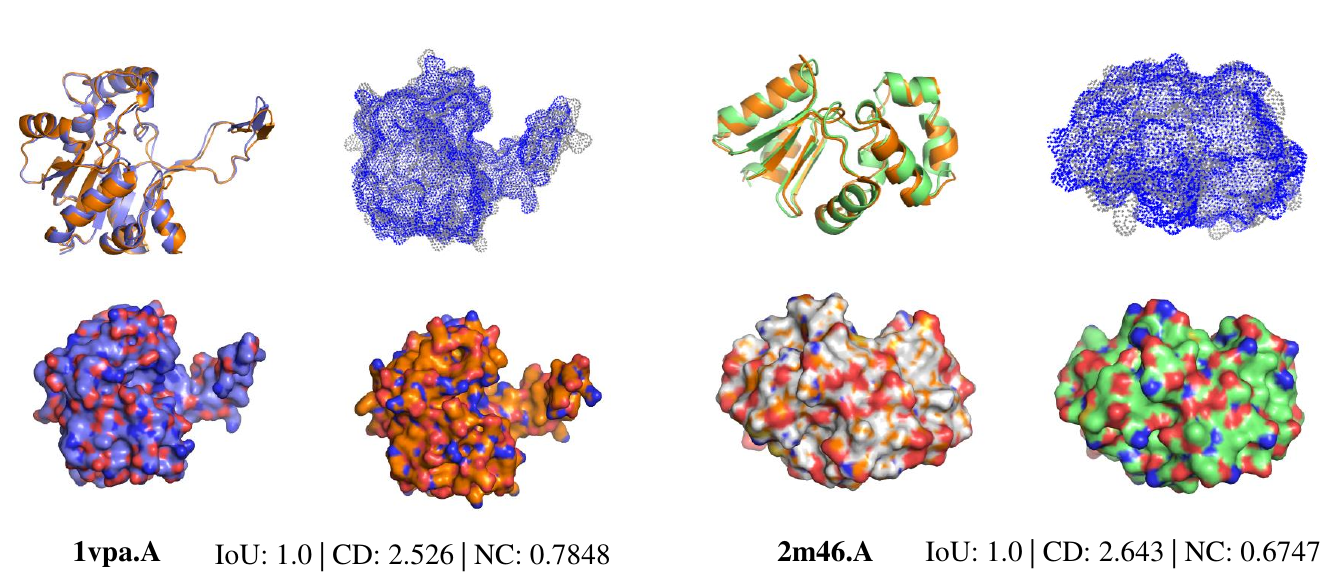}
     \caption{Comparison between original and designed surfaces, where molecular surfaces are visualized from two perspectives: the point cloud view and the manifold view. }
     \label{fig:surf_visual}
 \end{figure}

In addition to restoring protein structure, we quantify surface similarity between the designed and ground-truth proteins.  We envision the surfaces of the designed and target proteins in Fig.~\ref{fig:surf_visual}. A substantial overlap is observed between the point clouds of the designed protein surface and the ground-truth protein surface, with a relatively low CD and a significantly high IoU. All of these indicate that SurfDesign produces proteins with the expected surface shapes. 

\end{document}